\documentclass[conference]{IEEEtran}
\IEEEoverridecommandlockouts
\usepackage{cite}
\usepackage{amsmath,amssymb,amsfonts}
\usepackage{algorithmic}
\usepackage{graphicx}
\usepackage{textcomp}
\usepackage{xcolor}
\usepackage{comment}
\usepackage{hyperref}
\usepackage{booktabs}
\usepackage{multirow}
\usepackage{comment}
\usepackage{csquotes}
\usepackage{refcount}
\usepackage{enumitem}

\usepackage{url}
\def\BibTeX{{\rm B\kern-.05em{\sc i\kern-.025em b}\kern-.08em
    T\kern-.1667em\lower.7ex\hbox{E}\kern-.125emX}}

\usepackage{glossaries}
\newacronym{LLM}{LLM}{Large Language Model}

\newacronym{NVP}{NVP}{N-Version Programming}

\newacronym{PFD}{PFD}{Probability of Failure on Demand}
\newacronym{EL}{EL}{Eckhardt and Lee}
\newacronym{LM}{LM}{Littlewood and Miller}

\glsdisablehyper
\newcommand{\jrc}[1]{\textcolor{orange}{#1}}

\begin{document}

\title{Effectiveness of LLM-based Software Diversity for Reliability Improvement -- an Empirical Study\\[0.5ex]
%{\Large Practical Experience Report}
}

\begin{comment}
\author{\IEEEauthorblockN{1\textsuperscript{st} Given Name Surname}
\IEEEauthorblockA{\textit{dept. name of organization (of Aff.)} \\
\textit{name of organization (of Aff.)}\\
City, Country \\
email address or ORCID}
\and
\IEEEauthorblockN{2\textsuperscript{nd} Given Name Surname}
\IEEEauthorblockA{\textit{dept. name of organization (of Aff.)} \\
\textit{name of organization (of Aff.)}\\
City, Country \\
email address or ORCID}
\and
\IEEEauthorblockN{3\textsuperscript{rd} Given Name Surname}
\IEEEauthorblockA{\textit{dept. name of organization (of Aff.)} \\
\textit{name of organization (of Aff.)}\\
City, Country \\
email address or ORCID}
}
\end{comment}
\author{% 
    Gabriel Almeida\IEEEauthorrefmark{1}, 
    Ilir Gashi\IEEEauthorrefmark{2},
    Vladimir Stankovic\IEEEauthorrefmark{2}, 
    João R. Campos\IEEEauthorrefmark{1}\\[0.8ex] 
    \IEEEauthorblockA{%
        \makebox[\textwidth][c]{% 
            \begin{minipage}[t]{0.45\textwidth}
                \centering 
                \IEEEauthorrefmark{1}\textit{University of Coimbra, CISUC/LASI\\Department of Informatics Engineering} \\
                Coimbra, Portugal \\
                gabrielalmeida@student.dei.uc.pt,\\ jrcampos@dei.uc.pt
            \end{minipage}\hfill 
            \begin{minipage}[t]{0.45\textwidth}
                \centering 
                \IEEEauthorrefmark{2}\textit{Centre for Software Reliability\\Department of Computer Science\\City St George's, University of London} \\
                London, United Kingdom \\
                ilir.gashi.1@citystgeorges.ac.uk, vladimir.stankovic.1@citystgeorges.ac.uk
            \end{minipage}%
        }% 
    }% 
}

\maketitle

\begin{abstract}
Software diversity has been extensively studied as a means of reducing the risk of common-mode failures. Classic work showed that the central issue is whether failures of diversely redundant components  overlap in ways that limit the reliability gains. Traditional software diversity is costly to obtain, since it requires multiple implementations as well as the corresponding validation, maintenance, and deployment effort. Recent advances in \glspl{LLM} may change this. LLMs enable inexpensive code generation: they produce many candidate implementations of the same specification quickly, across different models, decoding settings, and programming languages. This raises a natural question: can \glspl{LLM} serve as practical generators of software diversity, and how much reliability improvement can that diversity actually provide? In this paper, we extend classical empirical studies of software diversity in human-written programs to \gls{LLM}-generated code. We study three specifications using both historical human-written programs and large pools of \gls{LLM}-generated ones evaluated under a common compilation, sandboxing, and exhaustive test suite. We explore \gls{LLM} diversity along multiple axes, including model family, generation temperature, and programming language. Reliability improvement is evaluated in a 1-out-of-2 configuration across both \textit{homogeneous} and \textit{heterogeneous} program populations, including within-\gls{LLM} pairings and pairings across programming languages and across \gls{LLM}-generated and human-written programs. The results show that combining \gls{LLM}-generated programs, especially in heterogeneous settings, can yield reliability gains, although this is partly conditioned by the programming language and generation setting. Taken together, these findings suggest that \glspl{LLM} provide a scalable source of comparatively low-cost programs whose diversity can be leveraged for reliability improvement.
\end{abstract}

\begin{IEEEkeywords}
Software Reliability, Diverse Redundancy, Empirical Evaluation
\end{IEEEkeywords}

%---------------------------------------------------------------------
\section{Introduction}
Software diversity has long been studied as a means of reducing the risk of common-mode failures in software-based systems. The underlying idea is central to \gls{NVP}: instead of relying on a single implementation, multiple independently developed programs of the same specification are combined in a fault-tolerant configuration, with the expectation that coincident failures will be sufficiently rare to improve overall dependability. This idea has been especially compelling in dependable and safety-critical systems, where the basic concepts of dependability emphasize reliability and safety in the presence of faults, and where redundancy is already a foundational architectural principle, from majority-voted schemes such as triple modular redundancy to fault-tolerant avionics and flight-control architectures that rely on replicated and, in some cases, diverse channels to preserve critical functionality under fault conditions~\cite{avizienis2004basic,lyons1962use,spitzer2018avionics}. In this context, \gls{NVP} can be understood as the software-level counterpart of established fault-tolerant design practice, aiming to reduce the impact of residual faults through multiple independently developed implementations of the same specification~\cite{avizienis1995methodology,eckhardt1985theoretical,littlewood1989conceptual,knight1986empirical,popov2014software}.

Classic work, however, showed that the effectiveness of such diversity cannot be reduced to a naive assumption of failure independence. \gls{EL} demonstrated that when two programs are drawn from the same development process the probability that both fail on the same test case is shaped by the distribution of test case difficulty across the test suite~\cite{eckhardt1985theoretical}. \gls{LM} generalized this view to heterogeneous developments, making explicit the role of covariance between the failure tendencies of different program populations~\cite{littlewood1989conceptual}. The well-known Knight and Leveson experiment provided early empirical evidence that independently developed programs generally do not fail independently~\cite{knight1986empirical}. However, this result is not absolute. As Cai \textit{et al.} later demonstrated, improvements in programmer training, stable specifications, and cleaner development protocols can drastically reduce the number of coincident failures, yielding empirical support for \gls{NVP}~\cite{cai2005experimental}. Subsequent work by van der Meulen and Revilla extended this by analyzing tens of thousands of human-written programs and quantifying the effectiveness of 1-out-of-2 diversity and of language diversity across large program populations~\cite{van2008effectiveness}.

Recent advances in \glspl{LLM} have shown remarkable potential for code generation. More importantly for this work, they enable inexpensive code generation across multiple programming languages and generation strategies. This creates a promising new empirical setting for software diversity research, since large numbers of candidate programs of the same specification can now be produced quickly and systematically. %The central question is therefore whether \glspl{LLM} can produce multiple solutions cheaply that differ in ways that reduce failure overlap and translate into measurable reliability improvement?	

Recent work explores \glspl{LLM} as generators of multiple programs, including automated \gls{NVP} construction~\cite{ron2024galapagos}, algorithmic diversity of model-generated code~\cite{lee2025diversely}, multi-language code-generation ensembles~\cite{xue2024multi}, and diversity -- quality trade-offs in \gls{LLM} outputs more broadly~\cite{shypula2025evaluating}. Yet these studies stop short of addressing whether, and to what extent, \gls{LLM}-generated programs exhibit failure diversity, and what is the magnitude of reliability improvement so achieved. 

We first replicate the human-written setting studied in prior work by reconstructing language-specific program pools from historical UVa Online Judge \footnote{\label{ft:uva}\url{https://onlinejudge.org/}} submissions for three specifications: \textit{3n+1}, \textit{Factors and Factorials}, and \textit{Factovisors}~\cite{van2008effectiveness,skiena2003programming}. We then generate large pools of \gls{LLM}-based programs for the same specifications, using multiple model families, programming languages, and temperatures.%, and iterative repair attempts, and evaluate all variants under a common compilation, sandboxing, and exhaustive benchmark pipeline.%  This enables three complementary analyses: reliability improvement within \gls{LLM}-generated pools, heterogeneous diversity across programming languages, and heterogeneous diversity across \gls{LLM}-generated and human-written solutions.

This paper makes three main contributions. {First}, to the best of our knowledge, it is the \textbf{first study to apply the classical \gls{EL} / \gls{LM} reliability-analysis framework to large populations of \gls{LLM}-generated code} and comparing it with prior studies on human-written programs. {Second}, it provides a \textbf{systematic characterization of \gls{LLM}-generated diversity} across multiple controllable dimensions and examines how these affect the reliability gains via diverse redundancy. {Third}, it \textbf{analyzes heterogeneous diversity between human-written and \gls{LLM}-generated programs}, offering an initial empirical view of whether, and to what extent, \glspl{LLM} reproduce existing failure patterns or instead provide complementary programs that improve reliability when paired with human-written code.

The human-written corpus used in this study comprises programs produced several years ago, providing a unique methodological advantage. By accessing a large population of programs developed before \glspl{LLM} became ubiquitous, we establish a contamination-free baseline that is not biased by AI-influenced coding patterns. This allows a rigorous evaluation of whether \gls{LLM}-generated diversity truly complements organic human problem-solving, offering a clear view of the fundamental heterogeneity between the two sources. While we acknowledge that this corpus is not representative of present-day \enquote{augmented} development, where human code is increasingly co-produced with AI, it remains valuable as a pre-\gls{LLM} reference baseline. This allows us to compare unaided human-written programs with AI-only generated programs under a common reliability framework, even though \glspl{LLM} are themselves trained on large corpora of human-written code. It serves as a controlled reference point for assessing whether \gls{LLM}-generated programs exhibit complementary failure behavior relative to unaided human-written code, and for grounding future studies of mixed human/\gls{LLM} development.

Our findings show that \gls{LLM}-generated programs can yield reliability gains, but these gains depend on the specification, programming language, and generation regime. More importantly, they suggest that \glspl{LLM} do not remove the classical challenge of correlated software failures; rather, they provide a scalable new source of heterogeneous programs whose diversity can be exploited for reliability improvement.

The remainder of this paper is organized as follows. Section~\ref{sec:bg} presents the background and related work, Section~\ref{sec:method} presents the methodology and Section~\ref{sec:results} presents the results. Section~\ref{sec:discussion} discusses the findings, Section~\ref{sec:threats} addresses threats to validity, and finally Section~\ref{sec:conclusion} concludes the paper and puts forward ideas for future work.
%--------------------------------------------------------------------- 
\section{Background and Related Work}
\label{sec:bg}

\subsection{Software Diversity for Reliability Improvement}

\begin{comment}
\begin{itemize}
    \item 	motivation for software diversity
	\item 	1-out-of-2 reliability
	\item 	Eckhardt–Lee and Littlewood–Miller
	\item 	Knight \& Leveson
	\item 	van der Meulen \& Revilla
	\item 	forced diversity / DSDs if brief and needed
\end{itemize}
\end{comment}

Software diversity is often used as a means of improving system reliability. Decision makers can use different forms of diversity in development, or selection, process and the associated mechanisms to reduce common failures between diversely redundant components.
We are interested in analyzing the effectiveness of diversity used for such purpose, for human-written and LLM-generated programs, and combinations thereof. This is best done by experimental evaluation.

Comprehensive experiments about diversity effectiveness were conducted several decades ago \cite{knight1986experimental,eckhardt1990experimental}.
They used relatively small samples: 27 and 20 versions, respectively, unlike our study. It would be (highly) desirable that failure independence between diverse programs holds in general. However, the well-known work in \cite{knight1986experimental} refuted that assumption – the experiments produced a counter-example. The validity of these experiments towards assessing diversity effectiveness in current development methods is somewhat limited.

\gls{EL} model \cite{eckhardt1985theoretical} explained why failure independence between diverse software cannot be assumed in general. A central concept is that of \textit{test case difficulty}: despite development teams possibly creating software versions that fail on different test cases, in developing their respective versions the teams find the same (subset of) test cases easy, and the same (subset of) test cases difficult. The EL model was generalized by \gls{LM} model \cite{littlewood1989conceptual} – the authors pointed out that each version may be developed by a different process, where different development teams use different approaches: e.g., different languages, different data structures and algorithms, etc. This is referred to as \textit{“forcing diversity”}. These conceptual, probabilistic models provided an important insight about use of diversity for fault tolerance: they explained that failure independence between diverse programs is just a possibility (probably unlikely), and that different levels of correlation (both positive or negative) between failure behavior of diverse software can be observed in practice, including the possibility that forced diversity (LM model) leads to a system that performs better than if the constituent programs failed independently. These models provide a theoretical basis for the work presented in this paper.

Experimental evidence of the effectiveness of forced diversity among different programs is scarce, especially for large populations of programs. Meine van der Meulen initiated a range of empirical studies \cite{van2008effectiveness}, and reported some initial results about the effect of a particular approach to forcing diversity – using diverse programming languages \cite{van2005effectiveness}. This was enhanced in \cite{popov2012empirical} where the authors evaluated the effectiveness of forced diversity using both diverse programming languages, and diverse program structure. They used a proxy measure of program structure based on two well-known software ‘complexity metrics’: Halstead Volume, and Cyclomatic Complexity. We have followed the basic methodology used in these works, with several extensions, most notably the one for analyzing diversity effectiveness of LLM-generated code.

\subsection{LLMs for Code Generation and Diversity}
\glspl{LLM} have become an increasingly important part of modern software engineering practice, supporting tasks such as code completion, code generation, translation, and debugging~\cite{peng2023impact,ziegler2022productivity,chen2024survey,zheng2025towards}. In particular, text-to-code generation has emerged as one of their most visible capabilities: given a natural-language description, modern models can produce executable implementations across a range of programming languages~\cite{chen2021evaluating,austin2021program,hendrycks2021apps,khan2023xcodeeval}. These capabilities are already influencing everyday development workflows, but they also make it possible to generate multiple candidate implementations of the same specification quickly and systematically~\cite{peng2023impact,ziegler2022productivity}.

However, \gls{LLM}-generated code remains error-prone. Prior work shows that generated programs may fail to compile, crash at runtime, omit relevant corner cases, hallucinate identifiers or APIs, or otherwise produce functionally incorrect programs~\cite{wang2025towards,tambon2025bugs,liu2024quality,chen2024survey}. Most current benchmark-style evaluations emphasize aggregate correctness, typically through execution-based metrics such as pass@k or unit-test success on datasets such as HumanEval, APPS, and xCodeEval~\cite{chen2021evaluating,hendrycks2021apps,khan2023xcodeeval,chen2024survey}. These evaluations are important, but they do not directly address to what extent multiple generated programs fail on the same test cases: the essential consideration when software diversity is used for reliability improvement.

This motivates a different view of \glspl{LLM}: not only as generators of individual programs, but as generators of diverse program populations. Diversity can be induced along several controllable axes, including model family, decoding strategy, target programming language. Recent studies suggest that some of these factors, especially temperature and the combination of heterogeneous models, can affect the diversity of generated code~\cite{lee2025diversely,xue2024multi,shypula2025evaluating}. However, the extent to which these sources of variation translate into useful diversity for reliability improvement remains largely unexplored. 

\subsection{Related Work}

A growing body of work has started to examine \glspl{LLM} as generators of multiple programs, but from perspectives that remain different from ours. Gal\'apagos, for example, investigates the automated construction of \gls{NVP} systems with \glspl{LLM}, focusing on the generation of functionally equivalent programs for protection against compiler miscompilation faults~\cite{ron2024galapagos}. While closely related, our focus is not limited to miscompilation faults. Instead, we evaluate software diversity as a broader mechanism for reliability improvement, through statistical analysis of failure overlap and of homogeneous and heterogeneous pairings across large program populations.

Other recent studies focus on diversity itself rather than on reliability improvement. Lee et al.~\cite{lee2025diversely} study the algorithmic diversity of model-generated code and show that temperature and heterogeneous models can increase diversity, but without evaluating the reliability gains from combining programs. Xue et al.~\cite{xue2024multi} propose a multi-language ensemble strategy for improving code-generation accuracy, again emphasizing correctness rather than software-diversity effectiveness. Shypula et al.~\cite{shypula2025evaluating} study diversity--quality trade-offs in \gls{LLM} outputs more broadly, while Eagal et al.~\cite{eagal2025analyzing} investigate the dependability of \glspl{LLM} in generating behaviorally equivalent programs. Taken together, these works show that \glspl{LLM} can generate multiple distinct code artifacts and that diversity can matter for correctness, robustness, or equivalence generation.

More generally, a large body of work evaluates \glspl{LLM} for code generation through correctness-focused benchmarks and error analyses~\cite{khan2023xcodeeval,chen2024survey,wang2025towards,tambon2025bugs}. These studies provide valuable insight into how well \glspl{LLM} solve programming tasks and what kinds of errors they make, but they do not analyze whether \gls{LLM}-generated programs provide useful software diversity in the classical reliability sense, namely by reducing failure overlap and improving the gains obtainable from redundancy.

%--------------------------------------------------------------------- 
\section{Methodology}
\label{sec:method}
Our methodology comprises three core stages: (1) compiling human-written and \gls{LLM}-generated program pools, (2) evaluating within-source reliability improvements (\textit{homogeneous} and language-forced), and (3) assessing cross-source (\textit{heterogeneous}) diversity between human and AI programs.

\subsection{Reliability Evaluation Framework}
We adopt the empirical framework from prior software-diversity studies to evaluate reliability improvement via redundant programs. We analyze a \textbf{1-out-of-2} fault-tolerant configuration, in which the system fails only if \textit{both} constituent programs fail on the same test case. We evaluate \textit{homogeneous} pairs (drawn from the same population) and \textit{heterogeneous} pairs (drawn from different populations, i.e., sub-pools).

Each program is evaluated against a common test suite for a given specification. We use a whole population of programs, as well as language-specific (C, C++, Java, and Pascal) sub-pools to analyze failure behavior.

\subsection{Specifications and Test Suites}
We selected three specifications from prior software-diversity studies:
\begin{itemize}[leftmargin=*]
    \item \textit{3n+1} (UVa 100)\footnote{\url{https://onlinejudge.org/external/1/100.pdf}}: Calculates the maximum cycle length for the 3n+1 algorithm between integers $i$ and $j$. We used 5,000 test cases with $i \in [1, 100]$ and $j \in [1, 50]$.
    \item \textit{Factovisors} (UVa 10139)\footnote{\url{https://onlinejudge.org/external/101/10139.pdf}}: Determines whether $n!$ is divisible by $m$. We used 5,000 test cases sampled from ranges matching the 3n+1 specification.
    \item \textit{Factors and Factorials} (UVa 160)\footnote{\url{https://onlinejudge.org/external/1/160.pdf}}: Outputs the frequency of each prime number in the factorization of $N!$ ($2 \le N \le 100$). We used the  entire input range of 99 test cases.
\end{itemize}

\subsection{Measuring Diversity Effectiveness}
We evaluate diversity effectiveness using the \textit{reliability improvement ratio}, $R = \frac{PFD_A}{PFD_{AB}}$ \cite{popov2012empirical,van2008effectiveness}, defined as the average \gls{PFD} of individual programs divided by the average \gls{PFD} of the resulting 1-out-of-2 pairs. Assuming all test cases and programs are equiprobable, $R$ serves as a standard metric to investigate potential reliability gains across varying marginal PFDs.

To analyze $R$ across different baseline PFD values, we simulate process improvement by filtering the pools based on a strict PFD threshold, incrementally discarding the least reliable programs \cite{popov2012empirical}. Because the pools contain a high concentration of \textit{Perfect} programs (\gls{PFD} = 0), the average \gls{PFD} of a filtered pool can be significantly lower than the minimum non-zero failure rate of a single program (e.g., $1/5000$ or $2 \cdot 10^{-4}$). As the threshold incrementally isolates highly reliable programs, this concentration of perfect solutions dilutes the pool's average \gls{PFD} into the $10^{-7}$ and $10^{-8}$ ranges. For heterogeneous analysis, drawn sub-pools must share matching average single-program PFDs before plotting the empirical $R$ across these sliding thresholds.

\subsection{Human-Written Programs} \label{sec:human_rep}
We use programs from UVa Online Judge repository written for the three specifications in one of the 4 programming langauges: C, C++, Java, and Pascal.

Programs were compiled and executed in a sandboxed environment to generate a boolean \textit{failure vector} (pass/fail outcomes across all test cases). Programs exhibiting identical failure vectors belong to the same \textit{score class} \cite{popov2012empirical}.

Following prior heuristics, we applied a 0.2-second timeout per test case (dynamically adjusted for Java's JVM startup). Additionally, we halted programs timing out on 30 consecutive test cases \cite{popov2012empirical}. We retained only the first valid program per author, excluding execution failures and completely incorrect programs (\gls{PFD}=1) to form the human baseline pools.

\subsection{LLM-Generated Programs}
To induce diversity, we utilized 14 modern \glspl{LLM}, five temperature settings ($[0.5, 0.75, 1.0, 1.25, 1.5]$), and the four target programming languages. The models comprised five Commercial APIs (\textit{gemini-3.1-pro-preview}, \textit{gemini-2.5-pro}, \textit{gpt-5-mini}, \textit{gpt-4o-mini}, \textit{gpt-5.1-codex-mini}) and nine Local open-weight models (\textit{Qwen2.5-Coder} [7B, 14B, 32B], \textit{DeepSeek-Coder-V2-Lite}, \textit{llama-3.3-70b-instruct-awq}, \textit{Codestral-22B-v0.1}, \textit{Mistral-Small-Instruct-2409}, \textit{Mixtral-8x7B-Instruct-v0.1}, and \textit{falcon-mamba-7b-instruct}). We requested 100 programs per configuration, yielding up to 2,000 programs per model (or 400 for APIs lacking temperature control).

The system prompt instructed the model to act as an expert programmer and output only the requested code inside a markdown block, without explanations or conversational text. The user prompt requested a program reading from standard input and printing to standard output, followed by the verbatim specification description from the UVa Online Judge.

All \gls{LLM}-generated programs faced the exact same compilation, sandboxing, and timeout constraints as the human baseline. Accepted programs were evaluated against the full test suites. A program was acceptable for inclusion upon succeeding in at least one test case. Consistent with human baseline filtering, completely incorrect programs (\gls{PFD}=1) were excluded to form the final \gls{LLM} diversity pools.

\subsection{Diversity Analysis}
We conducted a two-part analysis: within-source (evaluating human and \gls{LLM} pools separately) and cross-source (combining one human-written and one LLM-generated program).

\subsubsection{Within-Source Diversity Analysis}
For each source, we evaluated homogeneous and heterogeneous diversity under the \gls{EL} \cite{eckhardt1985theoretical} and \gls{LM} \cite{littlewood1989conceptual} models. Homogeneous 1-out-of-2 pairs were drawn from the same pool, whereas heterogeneous pairs combined different language-specific sub-pools. Applying this to the human-written corpus establishes a historical baseline and validates our pipeline. Replicating the analysis for \gls{LLM}-generated pools enables a direct, standardized comparison between human and AI failure diversity.

\subsubsection{Cross-Source Human--LLM Diversity}
We formed cross-source heterogeneous pairs to assess whether \glspl{LLM} introduce similar faults or act as a complementary diversity sources when combined with human-written programs. To interpret these system-level reliability results, we compared the programs' failure vectors, enabling granular, test-case-level mapping of failure overlap between human developers and \glspl{LLM}.

%--------------------------------------------------------------------- 
\section{Results}
\label{sec:results}

\subsection{Human-Written Software Diversity}
In this section, we present the baseline findings from the historical human-written dataset. We first characterize the distribution and composition of the human program pools for each specification. We then report the homogeneous 1-out-of-2 reliability improvement obtained from human-written programs, followed by the effects of programming-language diversity within the human pools.

\subsubsection{Human Program Pools}
To establish a baseline directly comparable with prior software diversity research, we first evaluated the raw historical UVa Online Judge programs. For each of the three specifications (\textit{3n+1}, \textit{Factors and Factorials}, and \textit{Factovisors}), programs were categorized into three groups based on their \gls{PFD}:
\begin{itemize}
    \item \textbf{Perfect (\boldmath{$PFD = 0$}):} Programs that successfully produced correct output for all test cases in the test suite.
    \item \textbf{Partial (\boldmath{$0 < PFD < 1$}):} Programs that produced the correct output for at least one test case, and failed on at least one other test case.
    \item \textbf{Incorrect (\boldmath{$PFD = 1$}):} Programs that failed to compile, crashed during execution, exceeded the timeout limit, or produced a wrong output for every test case in the test suite.
\end{itemize}

Table~\ref{tab:human_summary_all} presents the unified view of the raw program distribution and the final composition of the filtered diversity pools. Across all specifications, a significant portion of the initial raw programs fall into the \textit{Incorrect} category. For the reliability analyses, all \textit{Incorrect} programs ($PFD = 1$) were excluded, and only the first valid program per author was retained for each programming language. Valid programs can therefore be either \textit{Perfect} or \textit{Partial}. This filtering process drastically reduces the population size, but guarantees that the programs used for pairing are functionally viable and unique per submitter (e.g., the \textit{3n+1} pool is reduced from 95,659 raw programs to 15,295 valid programs).

\begin{table}[t]
    \centering
    \setlength{\tabcolsep}{3pt}
    \caption{Program Distribution and Filtered Diversity Pool for Human-Written Programs}
    % \resizebox{\columnwidth}{!}{%
    \begin{tabular}{llrrrrr}
        %\toprule
        & & \multicolumn{2}{c}{\textbf{Raw Programs}} & \multicolumn{3}{c}{\textbf{Filtered Diversity Pool}} \\
        \cmidrule(lr){3-4} \cmidrule(lr){5-7}
        \textbf{Spec.} & \textbf{Lang.} & \textbf{Total} & \textbf{Incorrect} & \textbf{Pool} & \textbf{Perfect} & \textbf{Partial} \\
        \midrule

        \multirow{5}{*}{\rotatebox[origin=c]{90}{\shortstack{\textit{3n+1}\\(5,000\\tests)}}}
        & C      & 42,421 & 22,446 &  5,205 & 2,431 & 2,774 \\
        & C++    & 27,628 &  6,903 &  5,754 & 2,853 & 2,901 \\
        & Java   & 17,910 &  2,134 &  2,593 & 1,822 &   771 \\
        & Pascal &  7,700 &  2,105 &  1,743 &   848 &   895 \\
        & \textbf{Global} & \textbf{95,659} & \textbf{33,588} & \textbf{15,295} & \textbf{7,954} & \textbf{7,341} \\
        \midrule

        \multirow{5}{*}{\rotatebox[origin=c]{90}{\shortstack{\textit{Fact. \& Fact.}\\(99\\tests)}}}
        & C      &  3,542 & 1,162 & 1,216 &   896 & 320 \\
        & C++    &  6,320 & 4,014 & 1,155 &   843 & 312 \\
        & Java   &    169 &    64 &    46 &    33 &  13 \\
        & Pascal &  1,127 &   318 &   423 &   319 & 104 \\
        & \textbf{Global} & \textbf{11,158} & \textbf{5,558} & \textbf{2,840} & \textbf{2,091} & \textbf{749} \\
        \midrule

        \multirow{5}{*}{\rotatebox[origin=c]{90}{\shortstack{\textit{Factovisors}\\(5,000\\tests)}}}
        & C      & 1,369 &   289 & 273 & 165 & 108 \\
        & C++    & 3,402 & 1,350 & 561 & 327 & 234 \\
        & Java   &   227 &    82 &  42 &  12 &  30 \\
        & Pascal &   211 &    45 &  59 &  43 &  16 \\
        & \textbf{Global} & \textbf{5,209} & \textbf{1,766} & \textbf{935} & \textbf{547} & \textbf{388} \\
        %\bottomrule
    \end{tabular}%
    %}
    \label{tab:human_summary_all}
\end{table}

\subsubsection{Homogeneous 1-out-of-2 Reliability Improvement}
\begin{figure*}[!t]
    \centering  
    % All Programs Homogeneous (Humans vs LLMs)
    \begin{minipage}{0.32\textwidth}
        \centering
        \includegraphics[width=\linewidth]{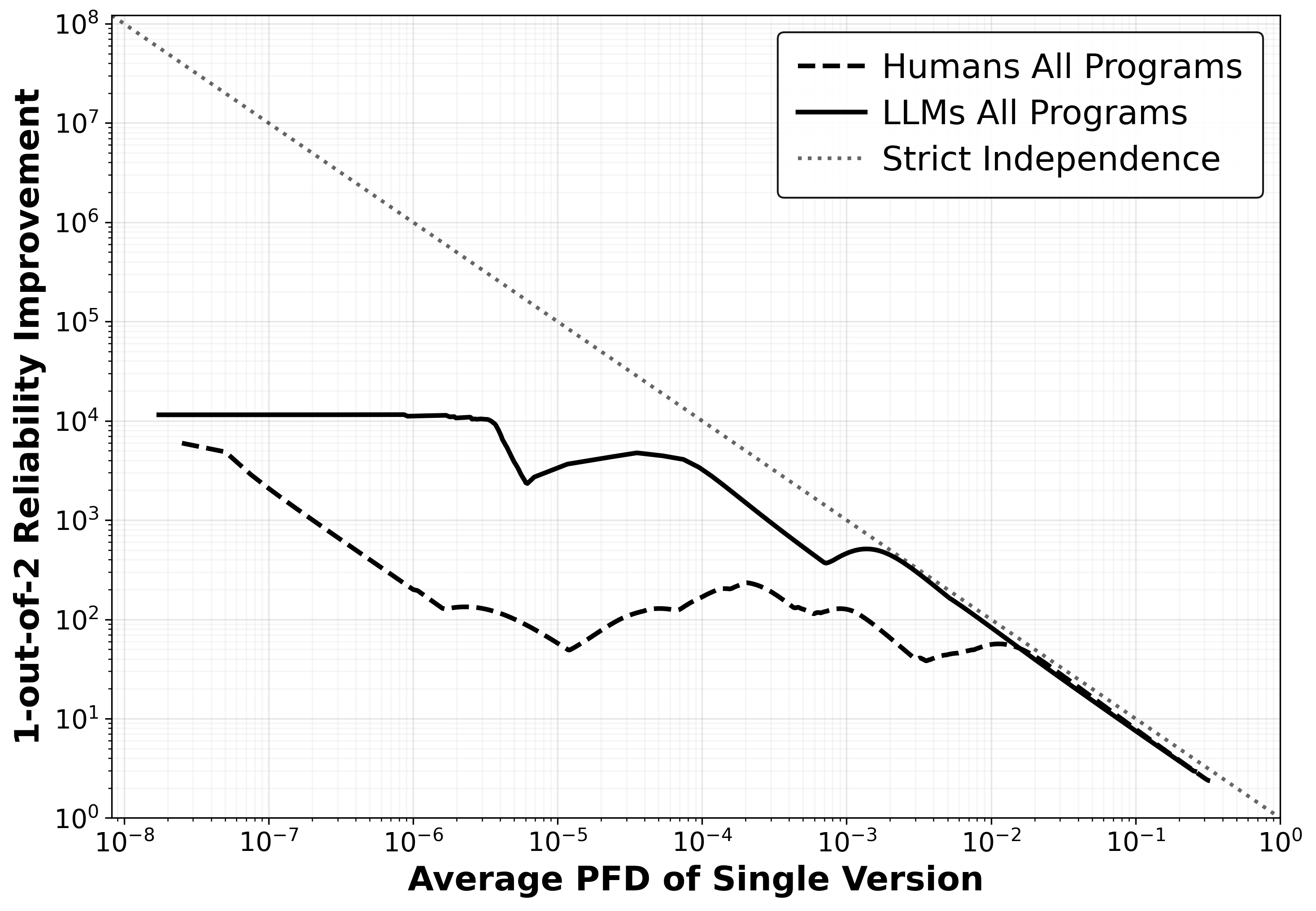}
        \centerline{(a) 3n+1 (All Programs)}
    \end{minipage}\hfill
    \begin{minipage}{0.32\textwidth}
        \centering
        \includegraphics[width=\linewidth]{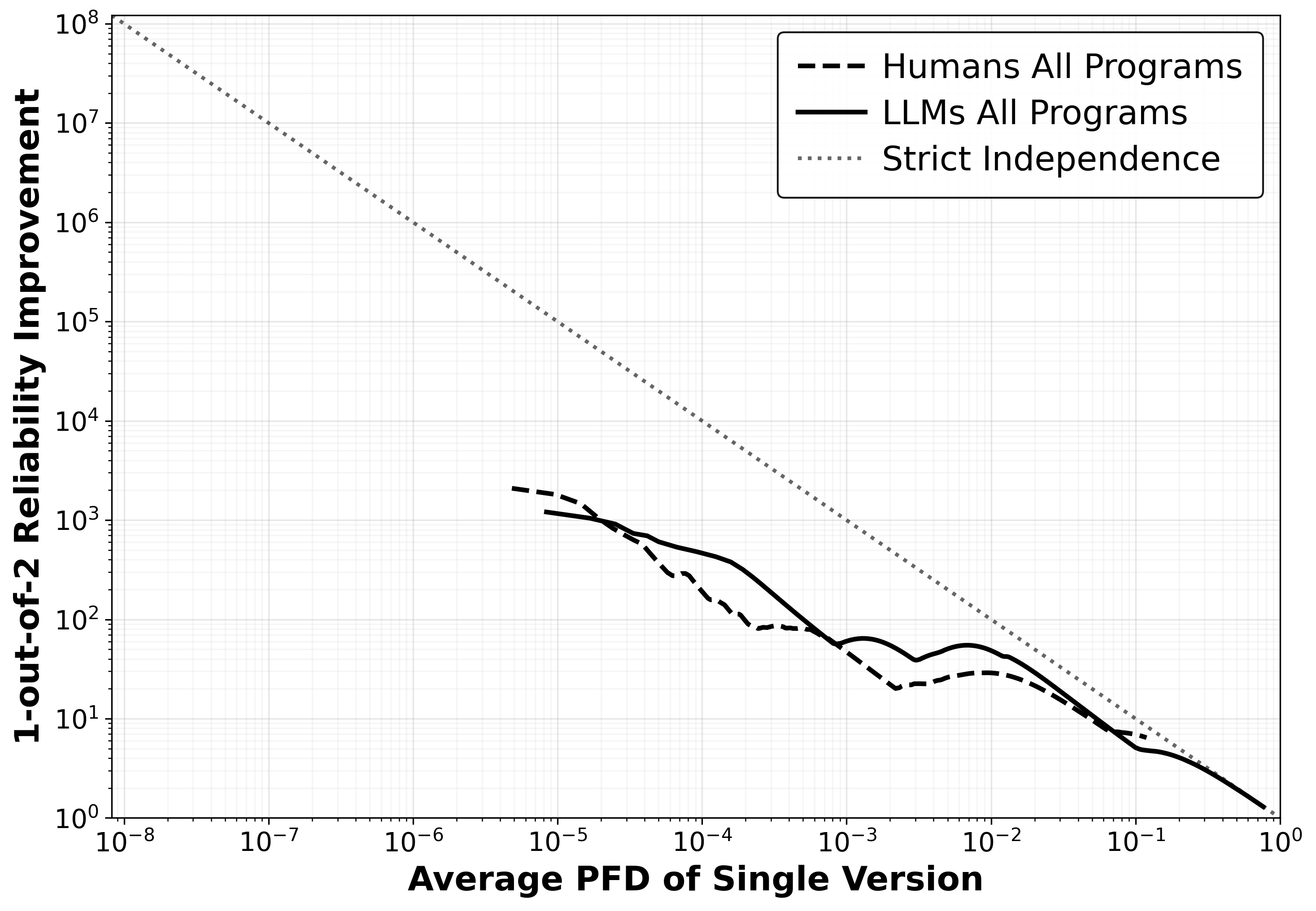}
        \centerline{(b) Factors \& Fact. (All Programs)}
    \end{minipage}\hfill
    \begin{minipage}{0.32\textwidth}
        \centering
        \includegraphics[width=\linewidth]{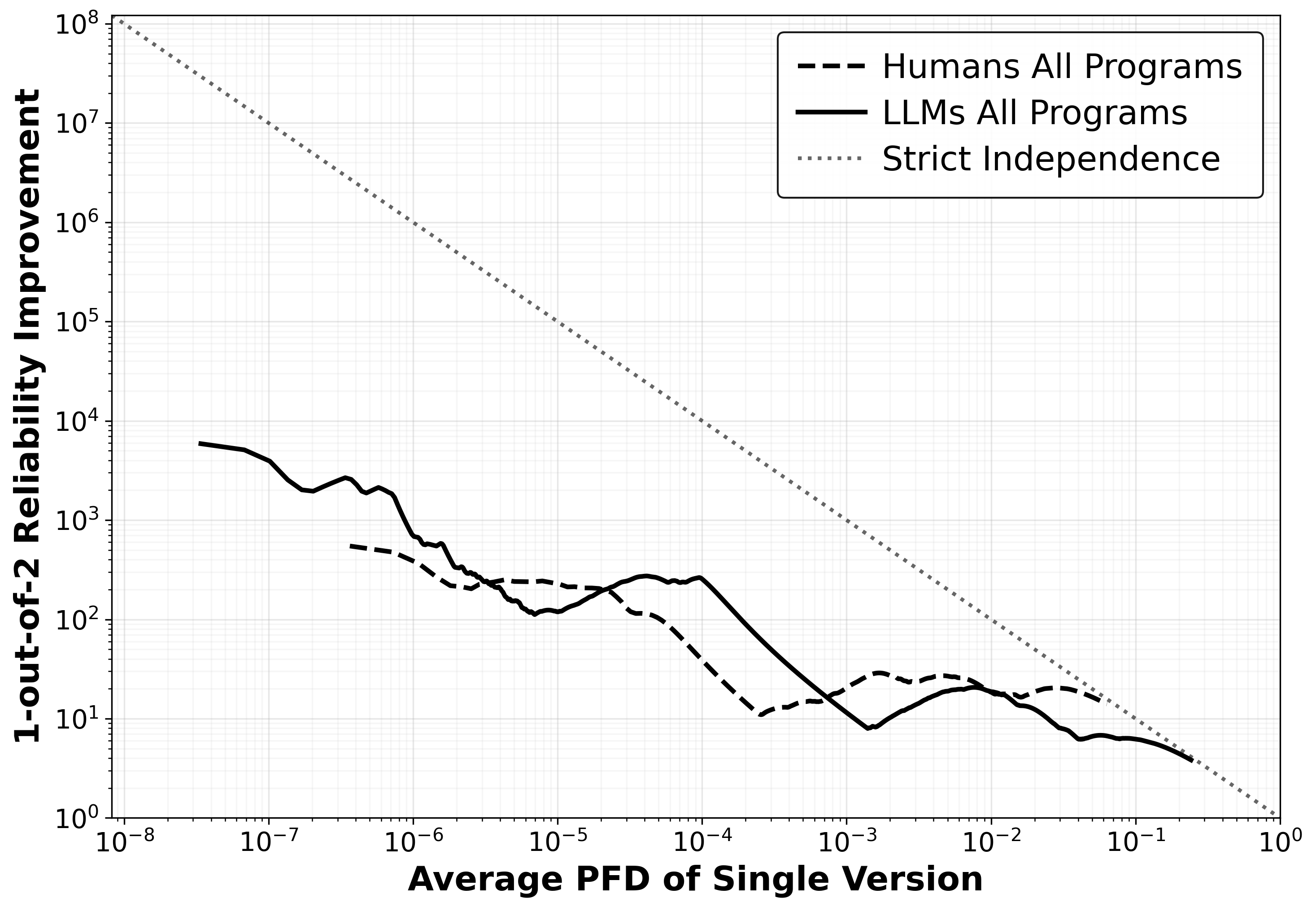}
        \centerline{(c) Factovisors (All Programs)}
    \end{minipage}
    \vspace{0.3cm}
    
    % Language-Specific Homogeneous (Humans vs LLMs)
    \begin{minipage}{0.32\textwidth}
        \centering
        \includegraphics[width=\linewidth]{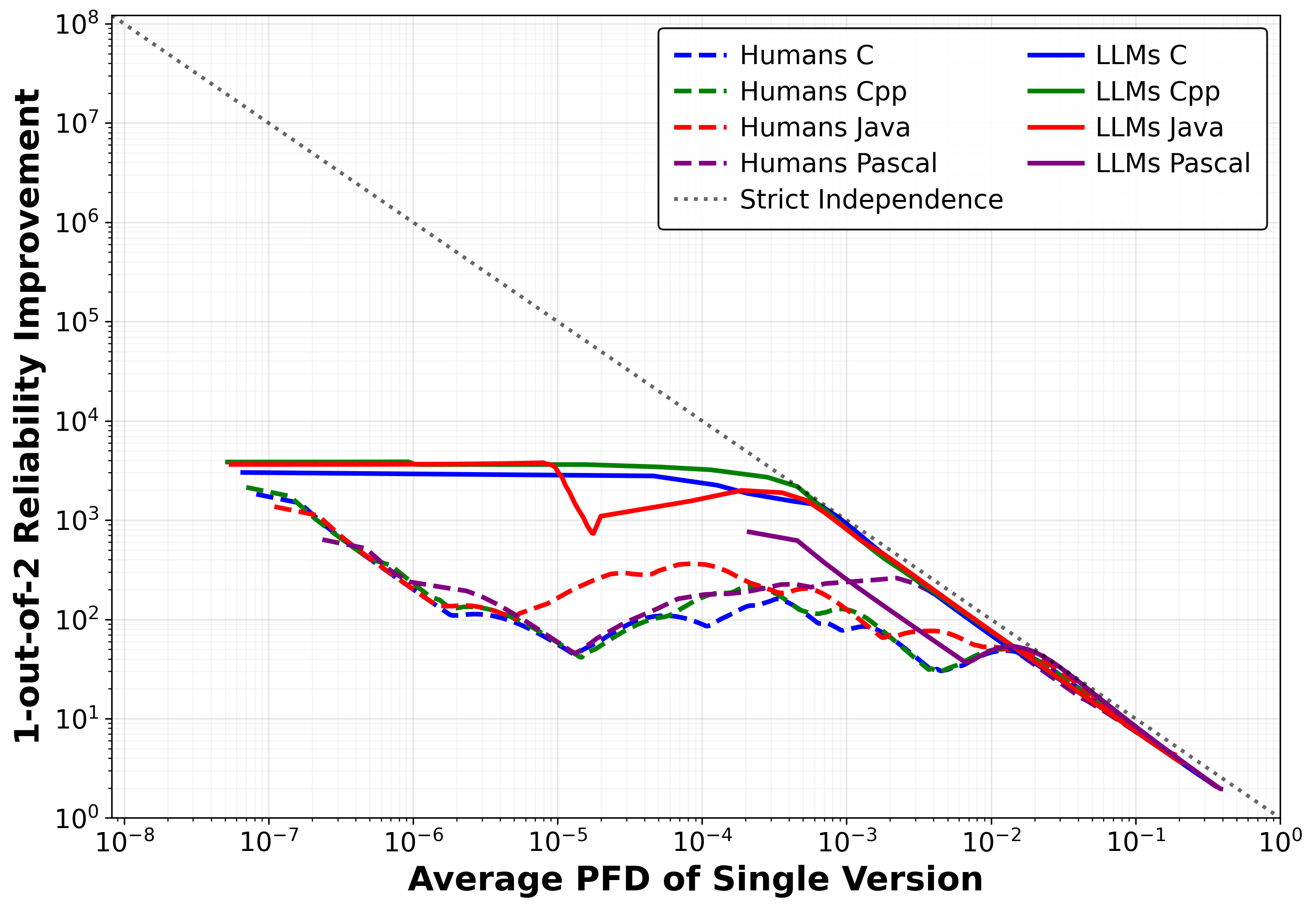}
        \centerline{(d) 3n+1 (By Language)}
    \end{minipage}\hfill
    \begin{minipage}{0.32\textwidth}
        \centering
        \includegraphics[width=\linewidth]{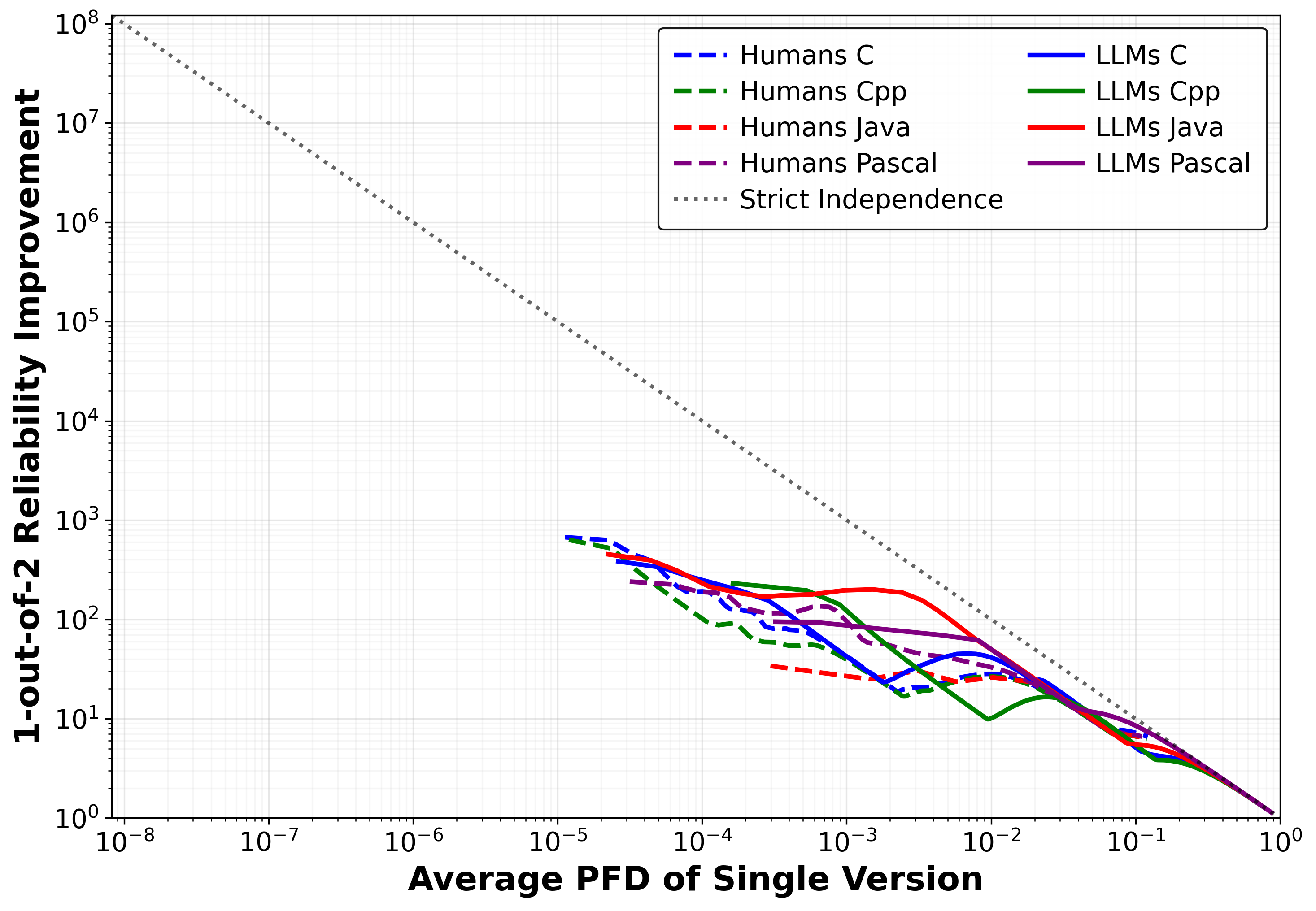}
        \centerline{(e) Factors \& Fact. (By Language)}
    \end{minipage}\hfill
    \begin{minipage}{0.32\textwidth}
        \centering
        \includegraphics[width=\linewidth]{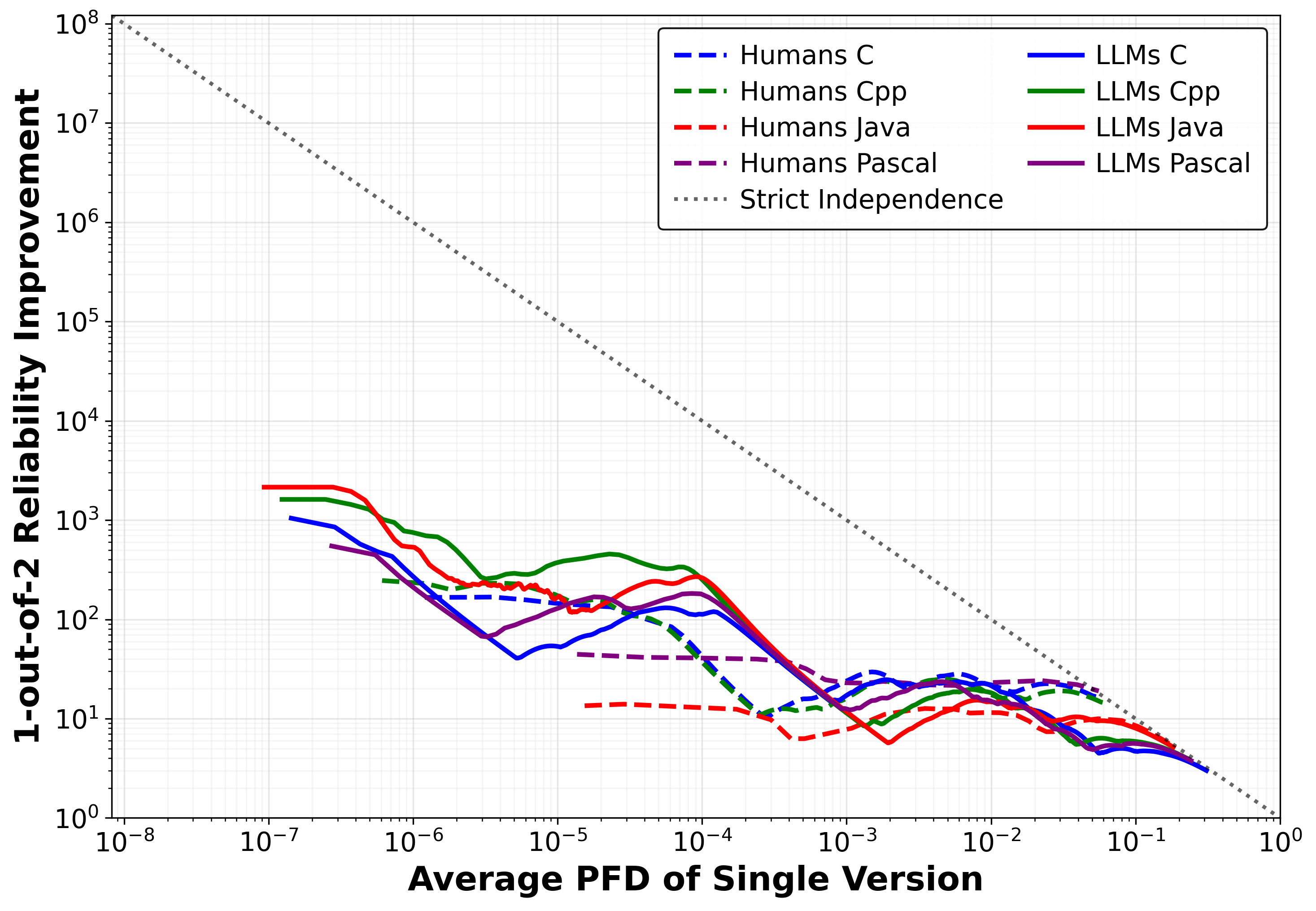}
        \centerline{(f) Factovisors (By Language)}
    \end{minipage}
    \caption{Homogeneous 1-out-of-2 reliability improvement comparing human-written and LLM-generated programs, analyzed as a unified pool of all programs (a-c) and by individual programming language (d-f).}
    \label{fig:homogeneous_combined_reliability}
\end{figure*}

To replicate the classical software-diversity analysis and establish the baseline for subsequent analyses, we evaluated the reliability improvement for 1-out-of-2 systems. Pairs were drawn from homogeneous pools of human-written programs, analyzing both a unified pool encompassing all programs (represented by the dashed lines in Figure \ref{fig:homogeneous_combined_reliability}a-c) and language-specific pools (dashed lines in Figure \ref{fig:homogeneous_combined_reliability}d-f).

Different programs tend to fail on the same subset of test cases, and thus their failures overlap, reflecting varying difficulty across the test suite. Consequently, the reliability improvement from diverse redundancy is lower than that predicted under hypotetical case of failure independence. These findings are consistent with the historical baseline of van der Meulen and Revilla \cite{van2008effectiveness}, demonstrating inherent failure correlation among human-written programs.

\subsubsection{Effects of Programming-Language Diversity}
\begin{figure*}[!t]
    \centering
    \begin{minipage}{0.32\textwidth}
        \centering
        \includegraphics[width=\linewidth]{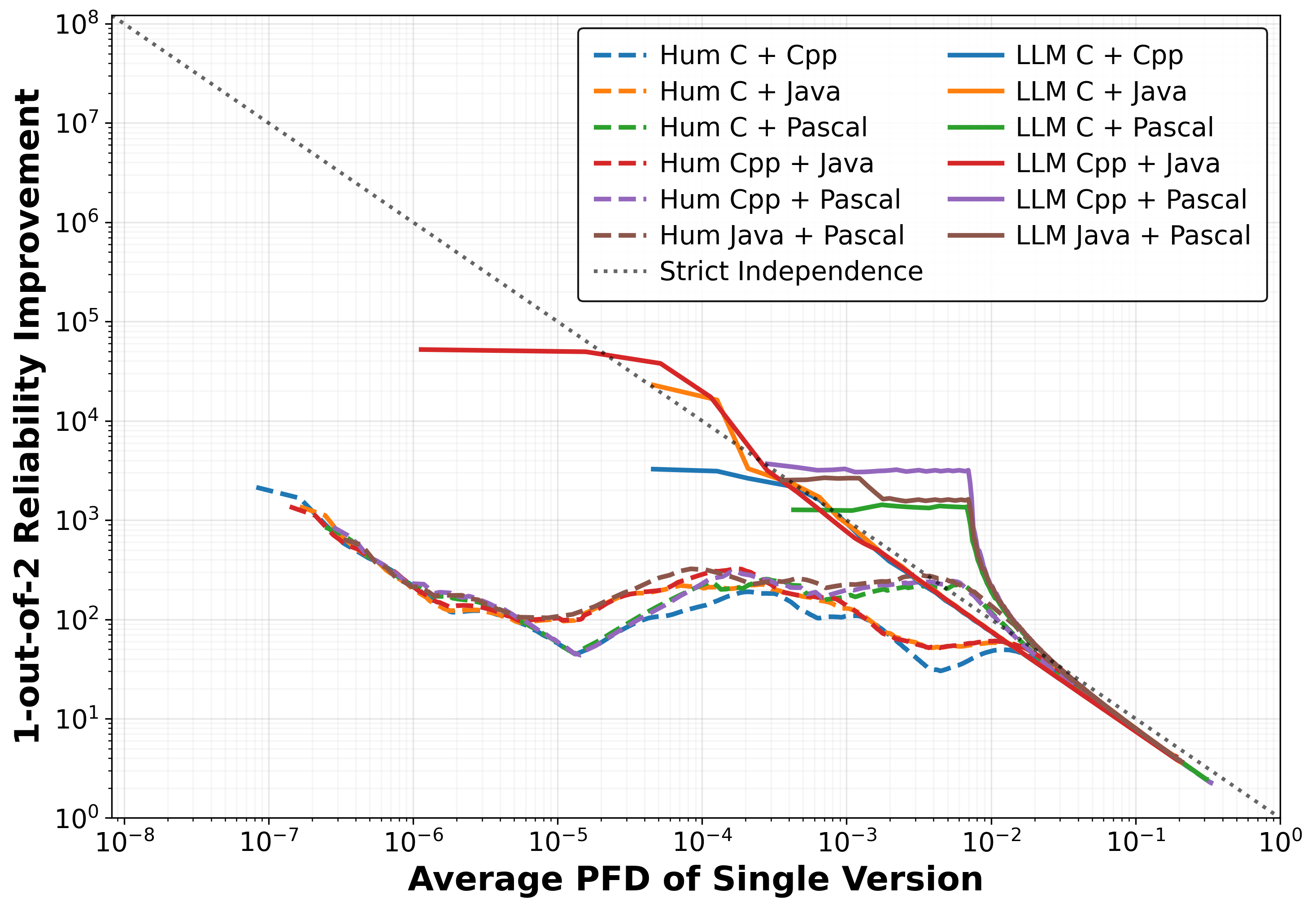}
        \centerline{(a) 3n+1}
    \end{minipage}\hfill
    \begin{minipage}{0.32\textwidth}
        \centering
        \includegraphics[width=\linewidth]{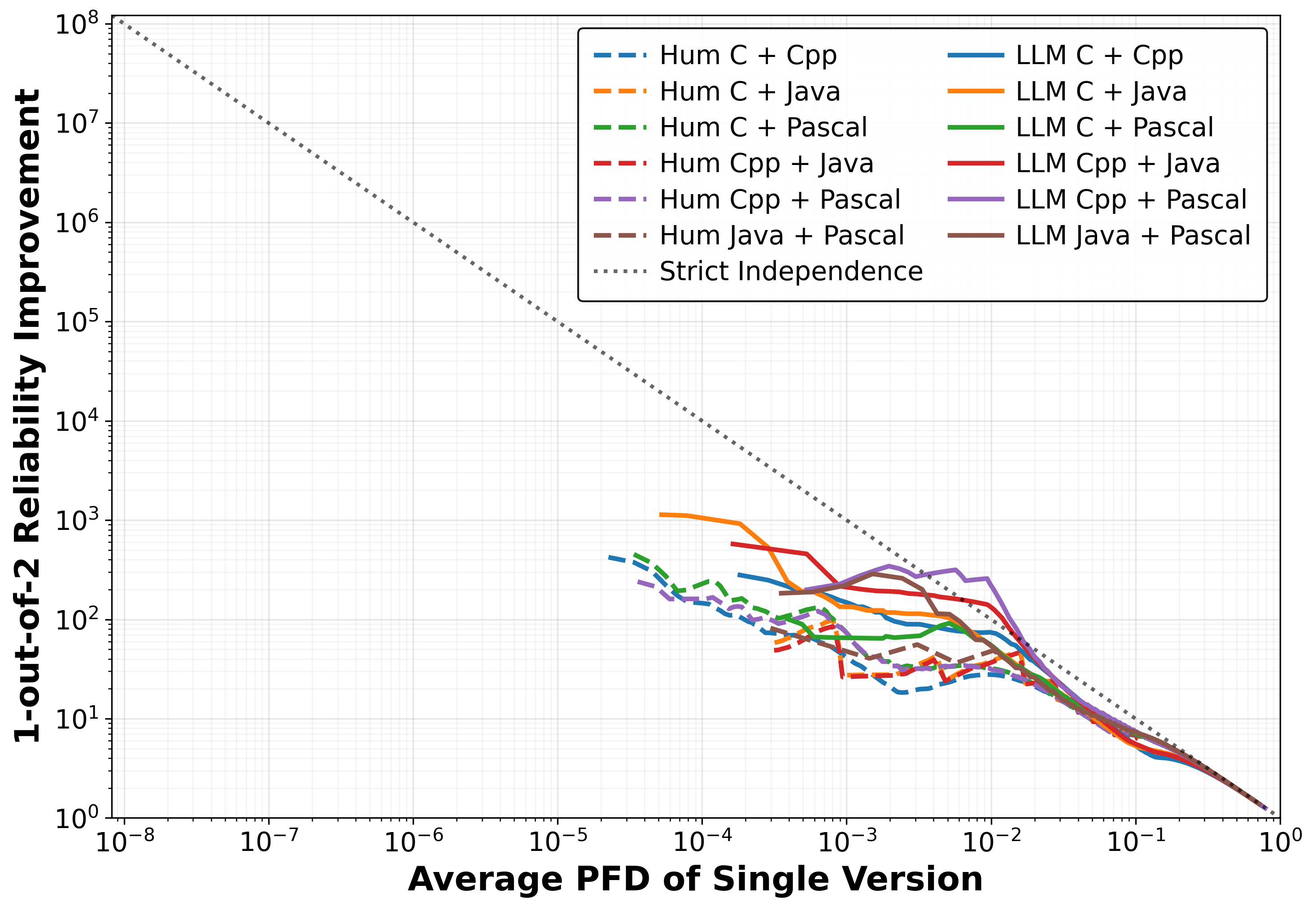}
        \centerline{(b) Factors and Factorials}
    \end{minipage}\hfill
    \begin{minipage}{0.32\textwidth}
        \centering
        \includegraphics[width=\linewidth]{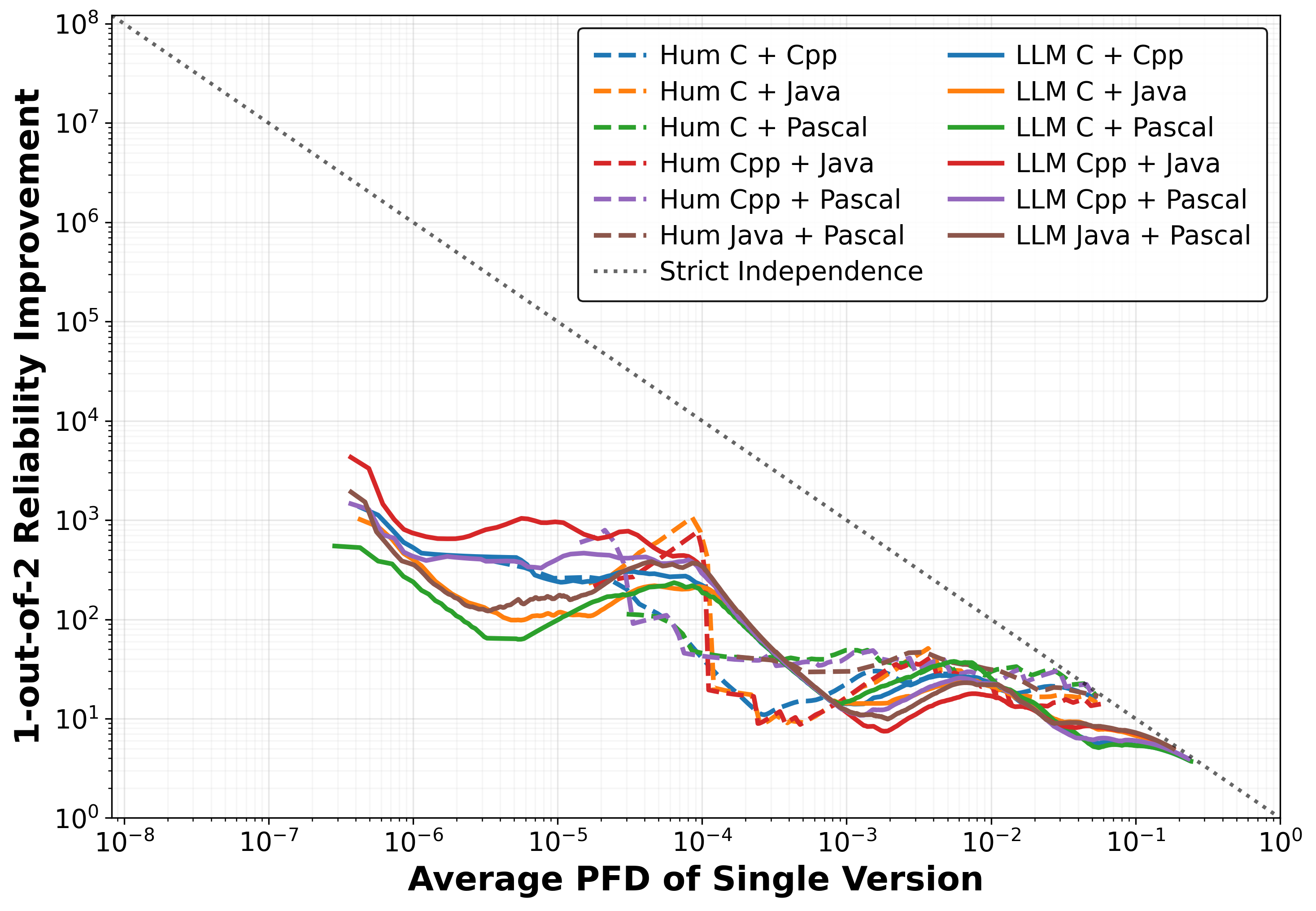}
        \centerline{(c) Factovisors}
    \end{minipage}
    
    \caption{Heterogeneous 1-out-of-2 reliability improvement across different programming languages, comparing human-written (dashed lines) and LLM-generated (solid lines) programs. 
    %\jrc{increase the fonts slightly, at least match fig}
    }
    \label{fig:heterogeneous_combined_reliability}
\end{figure*}

To investigate the impact of forced diversity through programming languages, we analyzed heterogeneous 1-out-of-2 pairings across the four languages (represented by the dashed lines in Figure \ref{fig:heterogeneous_combined_reliability}). This allows us to assess how distinct language ecosystems affect failure correlation.

The resulting empirical trends reveal a consistent hierarchy of diversity effectiveness. Pairings between closely related languages (e.g., C and C++) generally exhibit the lowest reliability gains, frequently mirroring their respective homogeneous baselines. In contrast, heterogeneous pairings using dissimilar languages (e.g., a C++ program paired with a Java or Pascal program) consistently yield higher reliability improvements, pulling the empirical measurements closer to the strict independence bound. Overall, forcing diversity through different programming languages results in higher reliability improvements compared to homogeneous pools. However, the improvement magnitude varies, and failure independence is not universally guaranteed, a conclusion consistent with prior empirical studies on forced diversity \cite{van2005effectiveness,van2008effectiveness}.

\subsection{\gls{LLM}-Generated Software Diversity}
In this section, we present the findings for the \gls{LLM}-generated corpus. We first characterize the distribution and composition of the \gls{LLM}-generated program pools, contrasting Local (Open-Weight) and Commercial API models. Next, we report the homogeneous 1-out-of-2 reliability improvement obtained from \gls{LLM}-generated programs, followed by the effects of programming-language diversity. Finally, we analyze cross-source heterogeneous pairings (Human vs. \gls{LLM}).

\subsubsection{\gls{LLM}-generated Program Pools}
We evaluated the raw \gls{LLM}-generated programs against the test suites. Table \ref{tab:llm_summary_all} details the global distribution of these initial programs across the three specifications. Consistent with the human baseline filtering criteria, we discarded all \textit{Incorrect} programs (those failing to compile or failing all test cases). 

\begin{table}[t]
    \centering
    \setlength{\tabcolsep}{3pt}
    \caption{\gls{LLM} Initial Program Distribution and Filtered Diversity Pool Composition}
    %\resizebox{\columnwidth}{!}{%
    \begin{tabular}{llrrrrr}
        %\toprule
        & & \multicolumn{2}{c}{\textbf{Raw Programs}} & \multicolumn{3}{c}{\textbf{Filtered Diversity Pool}} \\
        \cmidrule(lr){3-4} \cmidrule(lr){5-7}
        \textbf{Spec.} & \textbf{Tier \& Lang.} & \textbf{Total} & \textbf{Incorrect} & \textbf{Pool} & \textbf{Perfect} & \textbf{Partial} \\
        \midrule

        % ================= 3n+1 =================
        \multirow{12}{*}{\rotatebox[origin=c]{90}{\shortstack{\textit{3n+1} (5,000 tests)}}}
        & \multicolumn{6}{l}{\textbf{Local Models (Open-Weight)}} \\
        & \hspace{2mm} C      &  4,500 &  1,040 &  3,460 & 1,369 & 2,091 \\
        & \hspace{2mm} C++    &  4,500 &    926 &  3,574 & 2,155 & 1,419 \\
        & \hspace{2mm} Java   &  4,500 &    885 &  3,615 & 2,064 & 1,551 \\
        & \hspace{2mm} Pascal &  4,500 &  3,216 &  1,284 &   260 & 1,024 \\
        & \textbf{Local Global} & \textbf{18,000} & \textbf{6,067} & \textbf{11,933} & \textbf{5,848} & \textbf{6,085} \\
        \cmidrule{2-7}
        & \multicolumn{6}{l}{\textbf{Commercial Models (APIs)}} \\
        & \hspace{2mm} C      &  1,700 &     11  &  1,689 & 1,650 &    39 \\
        & \hspace{2mm} C++    &  1,700 &      2  &  1,698 & 1,693 &     5 \\
        & \hspace{2mm} Java   &  1,700 &     10  &  1,690 & 1,572 &   118 \\
        & \hspace{2mm} Pascal &  1,700 &    858 &    842 &   740 &   102 \\
        & \textbf{Com. Global} & \textbf{6,800} & \textbf{881} & \textbf{5,919} & \textbf{5,655} & \textbf{264} \\
        \midrule

        % ================= Factors and Factorials =================
        \multirow{12}{*}{\rotatebox[origin=c]{90}{\shortstack{\textit{Factors \& Factorials} (99 tests)}}}
        & \multicolumn{6}{l}{\textbf{Local Models (Open-Weight)}} \\
        & \hspace{2mm} C      &  4,500 &  3,627 &    873 &   277 &   596 \\
        & \hspace{2mm} C++    &  4,500 &  4,086 &    414 &    95 &   319 \\
        & \hspace{2mm} Java   &  4,500 &  3,724 &    776 &   301 &   475 \\
        & \hspace{2mm} Pascal &  4,500 &  4,422 &     78 &     9 &    69 \\
        & \textbf{Local Global} & \textbf{18,000} & \textbf{15,859} & \textbf{2,141} & \textbf{682} & \textbf{1,459} \\
        \cmidrule{2-7}
        & \multicolumn{6}{l}{\textbf{Commercial Models (APIs)}} \\
        & \hspace{2mm} C      &  1,700 &    757 &    943 &   106 &   837 \\
        & \hspace{2mm} C++    &  1,700 &    507 &  1,193 &   151 & 1,042 \\
        & \hspace{2mm} Java   &  1,700 &    605 &  1,095 &   150 &   945 \\
        & \hspace{2mm} Pascal &  1,700 &    687 &  1,013 &   116 &   897 \\
        & \textbf{Com. Global} & \textbf{6,800} & \textbf{2,556} & \textbf{4,244} & \textbf{523} & \textbf{3,721} \\
        \midrule

        % ================= Factovisors =================
        \multirow{12}{*}{\rotatebox[origin=c]{90}{\shortstack{\textit{Factovisors} (5,000 tests)}}}
        & \multicolumn{6}{l}{\textbf{Local Models (Open-Weight)}} \\
        & \hspace{2mm} C      &  4,500 &  1,092 &  3,408 &   243  & 3,165 \\
        & \hspace{2mm} C++    &  4,500 &  1,758 &  2,742 &   376 & 2,366 \\
        & \hspace{2mm} Java   &  4,500 &  1,133 &  3,367 &   999 & 2,368 \\
        & \hspace{2mm} Pascal &  4,500 &  3,136 &  1,364 &   152 & 1,212 \\
        & \textbf{Local Global} & \textbf{18,000} & \textbf{7,119} & \textbf{10,881} & \textbf{1,770} & \textbf{9,111} \\
        \cmidrule{2-7}
        & \multicolumn{6}{l}{\textbf{Commercial Models (APIs)}} \\
        & \hspace{2mm} C      &  1,700 &     48  &  1,652 & 1,150 &   502 \\
        & \hspace{2mm} C++    &  1,700 &    246 &  1,454 & 1,239 &   215 \\
        & \hspace{2mm} Java   &  1,700 &     96  &  1,604 & 1,147 &   457 \\
        & \hspace{2mm} Pascal &  1,700 &  1,028 &    672 &   578 &    94 \\
        & \textbf{Com. Global} & \textbf{6,800} & \textbf{1,418} & \textbf{5,382} & \textbf{4,114} & \textbf{1,268} \\
        %\bottomrule
    \end{tabular}%
%   }
    \label{tab:llm_summary_all}
\end{table}

The data reveals a distinct contrast in out-of-the-box reliability. For example, considering \textit{3n+1} specification, Commercial models achieved a 95.5\% (5,655 out of 5,919) \textit{Perfect} rate globally within their valid pools, alongside an \textit{Incorrect} rate of 13.0\% (881 out of 6,800) — lower than the historical human baseline (35.1\%). However, language selection heavily dictates success; while C, C++, and Java programs yielded high  rate of valid programs (84.5\%, or 15,726 out of 18,600 programs), Pascal proved highly problematic across both model tiers, registering a 65.7\% \textit{Incorrect} rate (4,074 out of 6,200 programs), likely reflecting scarce examples in pre-training corpora.

Disaggregating results by underlying architecture revealed several trends. Commercial APIs (notably the \textit{Gemini Pro} series) consistently achieved the highest proportion of \textit{Perfect} programs. Within the open-weight ecosystem, the \textit{Qwen2.5-Coder} family demonstrated competitive performance, exhibiting clear scaling laws where reliability improved sequentially from 7B to 32B parameter models. Conversely, models from the \textit{Mistral} family generated a low number of \textit{Perfect} programs but a notably high number of \textit{Partial} programs. Finally, architectures such as \textit{falcon-mamba-7b} proved ill-suited for algorithmic translation, yielding near 100\% \textit{Incorrect} rates.

The evaluation of the \textit{Factors and Factorials} specification revealed a heavily skewed distribution. While newer API models successfully generated highly reliable programs, a significant portion of the established models failed to produce a single \textit{Perfect} program. Instead, models such as \textit{gemini-2.5-pro} registered over 90\% \textit{Partial} programs, contributing significantly to the high concentration of \textit{Partial} programs in the All Programs pool for this specification.

\subsubsection{Homogeneous 1-out-of-2 Reliability Improvement}
In this homogeneous setup, pairs were drawn from the filtered initial generation pool, analyzing both a unified pool encompassing all programs (represented by the solid lines in Figure \ref{fig:homogeneous_combined_reliability}a-c) and language-specific pools (Figure \ref{fig:homogeneous_combined_reliability}d-f).

The curves for \gls{LLM}-generated programs also consistently fall below the theoretical expectation of failure independence (the dotted line). An analysis of the reliability improvements achieved by \glspl{LLM}, together with comparison against the human baselines, reveals several interesting insights. 

For \textit{3n+1} (Figure \ref{fig:homogeneous_combined_reliability}a, d), \gls{LLM}-generated programs significantly outperform human-written programs in the highly reliable spectrum. At a target \gls{PFD} of $10^{-7}$, the \textit{All Programs} \gls{LLM} pool achieves a reliability improvement of approximately 11,500x, compared to roughly 1,900x for the human pool. This trend is also noticeable in the language-specific analysis; within C++, for instance, homogeneous \gls{LLM} pairs yield a 3,850x improvement at $10^{-7}$, outperforming the 1,427x improvement of their human counterparts.
    
For \textit{Factors and Factorials} (Figure \ref{fig:homogeneous_combined_reliability}b, e), the results are less obvious. While the \gls{LLM} pool yields higher reliability improvements in the mid-range \gls{PFD} values ($10^{-2}$ to $10^{-4}$), the human pool demonstrates superior diversity at the extreme tail of reliability. At a \gls{PFD} of $10^{-5}$, the human \textit{All Programs} pool yields a 2,093x improvement, surpassing the 1,206x improvement of the \glspl{LLM}. The language breakdown reinforces this inversion: in C, humans achieve an 897x improvement at $10^{-5}$, more than double the \gls{LLM} equivalent (384x).
    
For \textit{Factovisors} (Figure~\ref{fig:homogeneous_combined_reliability}c, f), the reliability results are closely matched, with neither source showing a clear advantage. Within the \textit{All Programs} pool, humans outperform \glspl{LLM} at the $10^{-5}$ mark (yielding a 212x improvement versus 119x). However, the \glspl{LLM} rapidly achieve higher reliability improvements at $10^{-6}$ (657x vs 330x) and successfully generate programs capable of reaching individual \gls{PFD} values of $10^{-7}$ (achieving a 3,532x pair improvement). The language-specific breakdown reveals a human limitation in this specification: developers systematically failed to produce enough highly optimized programs to even register at the $10^{-7}$ threshold across any of the four individual languages, so the analysis at that level is effectively driven by the \gls{LLM}-generated pools. Notably, reliability estimates at these ultra-reliable extremes must be interpreted with caution; as pools deplete, a single residual fault type often becomes dominant, which can artificially inflate statistical gains \cite{van2008effectiveness}.

Ultimately, while \gls{LLM}-generated programs frequently achieve higher 1-out-of-2 reliability improvements than their human counterparts, their curves still follow a similar trajectory, remaining bounded by the independence line. This confirms that different \gls{LLM} programs, much like human programs, are inherently prone to overlapping failures.

\subsubsection{Effects of Programming-Language Diversity in \glspl{LLM}}
To investigate if forcing \glspl{LLM} to generate code in different programming languages induces failure diversity, we evaluated heterogeneous 1-out-of-2 pairings across all three specifications. Applying the \gls{LM} model \cite{littlewood1989conceptual}, we plotted all six unique combinations achievable with the four languages, shown in in Figure \ref{fig:heterogeneous_combined_reliability}. This allows us to observe how distinct language paradigms affect failure correlation in AI-generated programs.

A methodological point is needed to interpret these graphs. Unlike the historical human baseline, where curves extend more smoothly toward lower average \gls{PFD} values, the heterogeneous \gls{LLM} curves often begin abruptly at different points on the x-axis, and some are absent altogether. This occurs because heterogeneous pairing requires the compared language-specific pools to overlap in average \gls{PFD}. If one pool does not contain enough sufficiently reliable programs, then no comparison can be made in that region. This is particularly evident for Pascal, whose \gls{LLM}-generated pools contain very few high-quality programs and therefore fail to overlap with other languages at the lowest \gls{PFD} levels. As a result, Pascal-related heterogeneous curves are truncated or missing in the parts of the plots with lowest PFD values.

However, where data permits pairing, typically between C, C++, and Java, the results are highly encouraging. Comparing the solid AI curves against dashed human baselines, a direct numerical analysis of the empirical data reveals how \gls{LLM} language diversity often outperforms human language diversity.

For the \textit{3n+1} (Figure \ref{fig:heterogeneous_combined_reliability}a) specification, combining distinct ecosystems in \gls{LLM} prompts yields interesting reliability gains. When pairing C++ and Java \gls{LLM} programs at a target \gls{PFD} of $5 \cdot 10^{-5}$, the heterogeneous system achieves a 44,378x improvement. This significantly surpasses the equivalent human C++ and Java pairing, which manages only a 192.8x improvement at the same reliability threshold.
    
The \textit{Factors and Factorials} (Figure \ref{fig:heterogeneous_combined_reliability}b) specification showcases a mixed landscape. \glspl{LLM} excel in crossing major paradigm boundaries; for example, an \gls{LLM} C and Java pairing reaches a 1,362x improvement at $10^{-4}$, a threshold where human developers failed to produce enough valid programs to form an equivalent pair. Conversely, human heterogeneous pairs show greater resilience in closely related languages: human C and C++ pairs reach a 282.3x improvement at $5 \cdot 10^{-5}$, whereas the \gls{LLM} equivalent drops out of the chart entirely at that reliability level.
    
\textit{Factovisors} (Figure \ref{fig:heterogeneous_combined_reliability}c) demonstrates the \glspl{LLM}' ability to push further into the highly reliable spectrum through language forcing. In the C++ and Java pairing, humans achieve a strong 816.7x improvement at $10^{-4}$ (beating the \glspl{LLM}' 349.4x). However, the \gls{LLM} pairing curve extends all the way down to $5 \cdot 10^{-7}$ (achieving a 2,151x improvement), a deeply reliable space where the human pool again lacks sufficient valid programs to establish a baseline.

Empirical evidence demonstrates that programming language constraints successfully force \glspl{LLM} to navigate the algorithmic specification space differently. Despite the risk of early truncation in less-represented languages, pairing dominant languages mitigates common-mode failures effectively, frequently pushing the curves closer to the strict independence bound than human language diversity could achieve.

\subsection{Human--\gls{LLM} Heterogeneous Diversity}
\begin{figure*}[!t]
    \centering
    \begin{minipage}{0.32\textwidth}
        \centering
        \includegraphics[width=\linewidth]{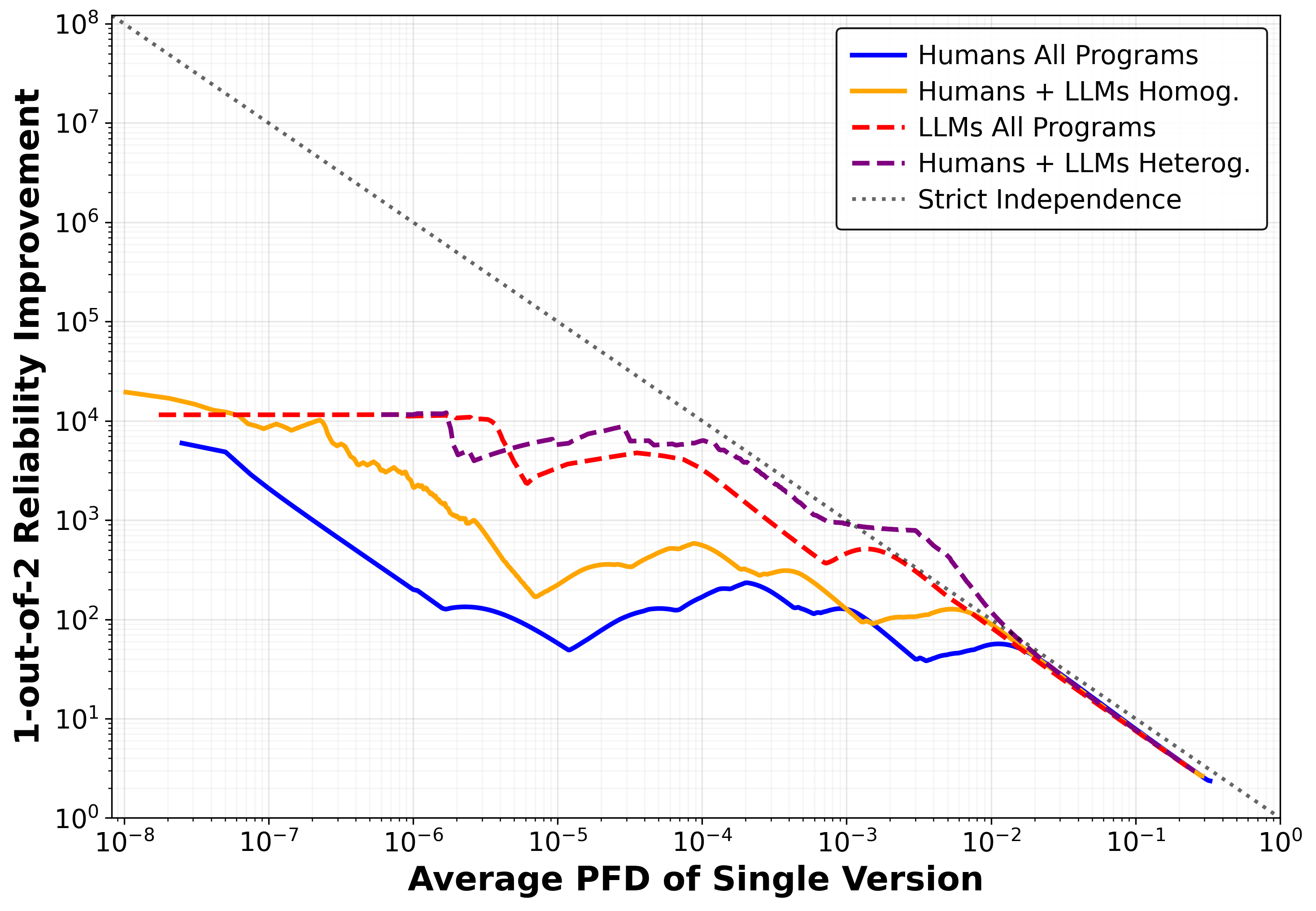}
        \centerline{(a) 3n+1}
    \end{minipage}\hfill
    \begin{minipage}{0.32\textwidth}
        \centering
        \includegraphics[width=\linewidth]{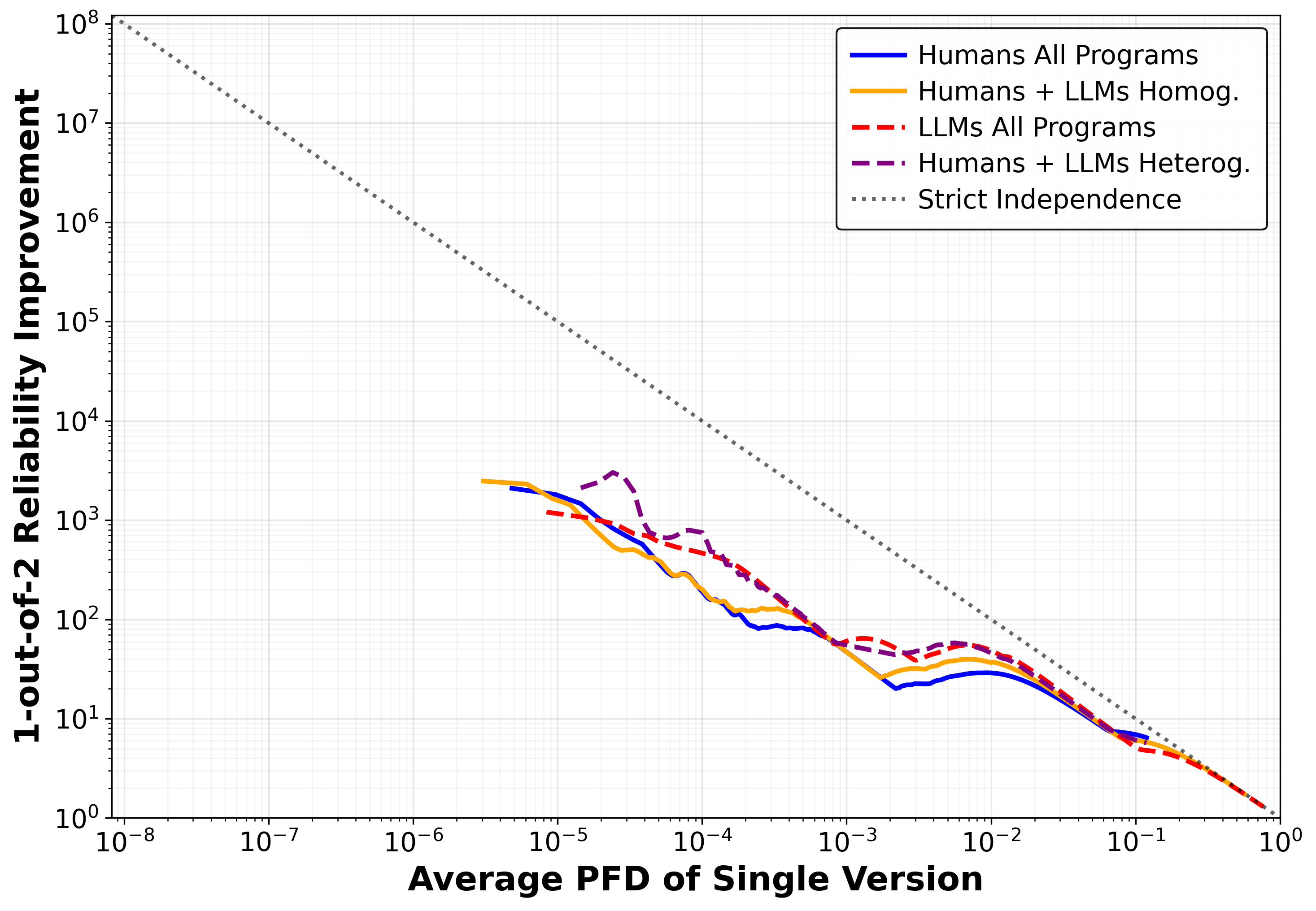}
        \centerline{(b) Factors and Factorials}
    \end{minipage}\hfill
    \begin{minipage}{0.32\textwidth}
        \centering
        \includegraphics[width=\linewidth]{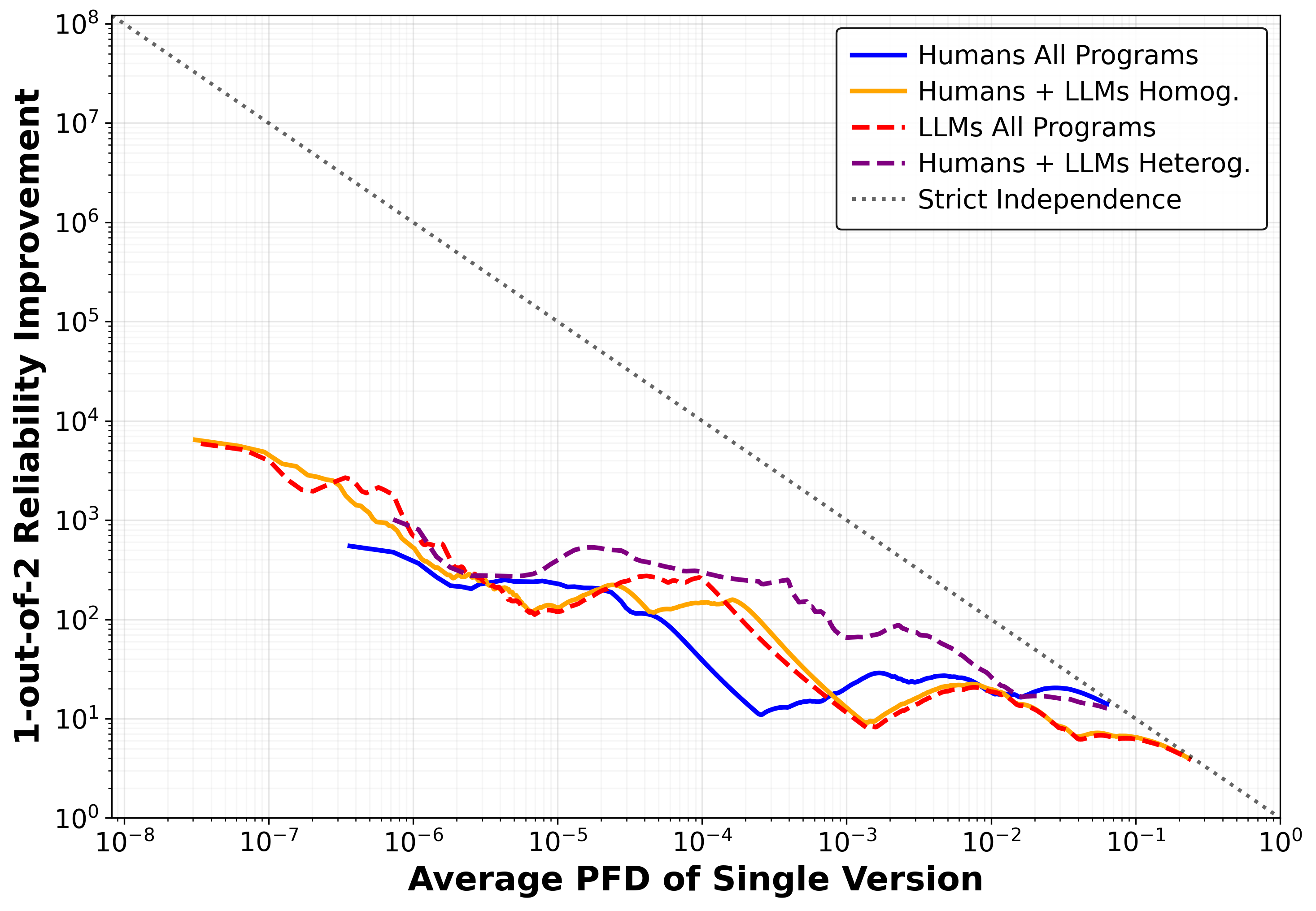}
        \centerline{(c) Factovisors}
    \end{minipage}
    \caption{Global 1-out-of-2 reliability improvement comparing pure configurations (\textit{Humans All Programs}, \textit{\glspl{LLM} All Programs}) against a homogeneous mixed pool (Eckhardt-Lee) \cite{eckhardt1985theoretical} and an enforced heterogeneous pairing (Littlewood-Miller) \cite{littlewood1989conceptual}.}
    \label{fig:human_llm_diversity}
\end{figure*}

This section addresses a core empirical question of our study: how does the failure diversity of pure \gls{LLM}-generated programs compare and interact with that of historically human-written programs? To answer this, we evaluate the reliability improvement of 1-out-of-2 systems utilizing different architectural pooling strategies: pure homogeneous pools, a unified mixed pool, and strictly enforced heterogeneous pairs. Figure \ref{fig:human_llm_diversity} consolidates these comparisons across the three specifications, plotting four distinct empirical scenarios.

To contextualize the effectiveness of cross-source collaboration, we first observe the pure, homogeneous baselines: systems built exclusively from human programs (solid blue line) and systems built exclusively from \gls{LLM} programs (dashed red line). The empirical data reveals a compelling hierarchy of reliability. Across the \textit{3n+1} and \textit{Factovisors} specifications, the homogeneous human baseline generally exhibits the highest degree of failure correlation, thereby yielding the lowest reliability improvement. For instance, at a target \gls{PFD} of $10^{-4}$ in \textit{3n+1} (Figure \ref{fig:human_llm_diversity}a), the human baseline yields a 168.1x improvement, while the pure \gls{LLM} baseline achieves 3,344.5x. However, this \gls{LLM} advantage is not absolute; in the \textit{Factors and Factorials} specification, the pure human baseline strongly outperforms the \gls{LLM} baseline when focusing on the highly reliable regions of the test suite. At a \gls{PFD} of $10^{-5}$ (Figure \ref{fig:human_llm_diversity}b), the pure human baseline reaches a 2,093.0x improvement compared to the \glspl{LLM}' 1,206.0x.

Beyond pure baselines, it is vital to evaluate the architectural approach to combining these two distinct development paradigms. Assuming \gls{EL} \cite{eckhardt1985theoretical} scenario, human and \gls{LLM} programs are placed into a single, massive resource pool, and two programs are drawn at random (represented by the solid orange \textit{Humans + \glspl{LLM} Homog.} curve). As shown in the plots, this homogeneous mix generally improves upon the pure human baseline by injecting \gls{LLM} diversity into the probability distribution (e.g., reaching 560.6x at $10^{-4}$ in \textit{3n+1}), but it often falls significantly short of the pure \gls{LLM} baseline by diluting the pool with highly correlated human failures.

However, the most substantial reliability gains are achieved by an \gls{LM} \cite{littlewood1989conceptual} architecture enforcing strict separation. In this scenario, the system ensures the pair is always heterogeneous: one component is human-written, and the other is strictly \gls{LLM}-generated (represented by the dashed purple \textit{Humans + \glspl{LLM} Heterog.} curve). Across almost all scenarios, this enforced pairing yields substantially higher reliability improvements than either pure or mixed homogeneous approaches, demonstrating stronger mitigation of coincident failures.

Empirical data demonstrates this forced pairing is superior to either pure source alone in avoiding coincident failures. In the \textit{Factovisors} specification at $10^{-5}$ (Figure \ref{fig:human_llm_diversity}c), the enforced Human-\gls{LLM} pair yields a 407.1x improvement, almost double the human baseline (212.0x) and more than triple the pure \gls{LLM} baseline (119.3x). For the \textit{3n+1} specification at $10^{-4}$, the heterogeneous pair achieves an extraordinary 6,342.3x improvement, effectively doubling pure \gls{LLM} performance.

In certain pools of this specification, the enforced heterogeneous curve actually crosses above the theoretical independence line. This reveals a rare instance of \textit{negative failure correlation}. Negative correlation means the sources possess actively complementary fault profiles. In these pools, if a human program stumbles on a specific test case, the \gls{LLM} program is \textit{less} likely to fail on that exact same test case. They actively compensate for each other's respective weaknesses, allowing the 1-out-of-2 pair to perform better than if the programs failed  independently.

Finally, disaggregating the \gls{LLM} pool reveals distinct behaviors between model tiers. Paired with humans at a target \gls{PFD} of $10^{-5}$, Commercial models provided robust heterogeneous improvements (e.g., 1,251.1x in \textit{Factovisors}), whereas Local models yielded much lower improvements (167.0x). Conversely, in \textit{3n+1}, the Human-Local pairing produced extreme mathematical anomalies, reaching an improvement factor over 2.9 million at $10^{-5}$, highlighting extremely low failure correlation and extreme variance in the individual baseline capabilities of the paired components.

\subsection{Explaining Diversity Effects}
While previous sections established reliability improvements across configurations, examining the underlying causes of coincident failures is necessary. We analyze the specification space at the test case level, including failure rates and difficulty landscapes, the overlap of human and \gls{LLM} failure profiles, and language-specific effects influencing aggregate conclusions. We also examine the impact of generation parameters, namely temperature effects on validity, average \gls{PFD}, and diversity improvement. %, as well as the role of iterative self-correction attempts in shaping diversity and reliability.

\subsubsection{Test Case-Level Failure Rates and Difficulty Landscapes}
To identify specific fault topologies, the analysis shifts from program-level reliability to test case-level difficulty. We calculated empirical failure rates for each test case across valid pools of human-written, Commercial, and Local programs. Test cases were classified into three groups based on failure rates: Low ($\le 5\%$), Medium ($5\% - 20\%$), and Hard ($> 20\%$).

The empirical data reveals that difficulty landscapes vary significantly by specification and source. In the \textit{3n+1} specification, both human-written programs and Local models found 3,725 of the 5,000 test cases to be Hard, whereas Commercial models processed the entire test suite as Low difficulty. For \textit{Factors and Factorials}, an inverse trend emerged: human-written programs predominantly faced Medium (81) or Low (14) difficulty across the test suite, while Commercial and Local programs systematically failed (categorizing 98 and 99 test cases as Hard, respectively). This systemic failure across the \gls{LLM} programs, contrasting with human success, indicates the models' inability to satisfy specific constraints of the test cases (such as exact output formatting expected by the oracle) regardless of their core algorithmic logic. Finally, in \textit{Factovisors}, human-written programs mostly encountered Medium difficulty (3,926 test cases). Commercial models proved robust, shifting most of the test suite into the Low bracket (4,128 test cases), whereas Local models struggled severely, registering 4,982 Hard test cases.

\subsubsection{Failure Profiles Overlap}
\begin{figure*}[!t]
    \centering
    \begin{minipage}{0.32\textwidth}
        \centering
        \includegraphics[width=\linewidth]{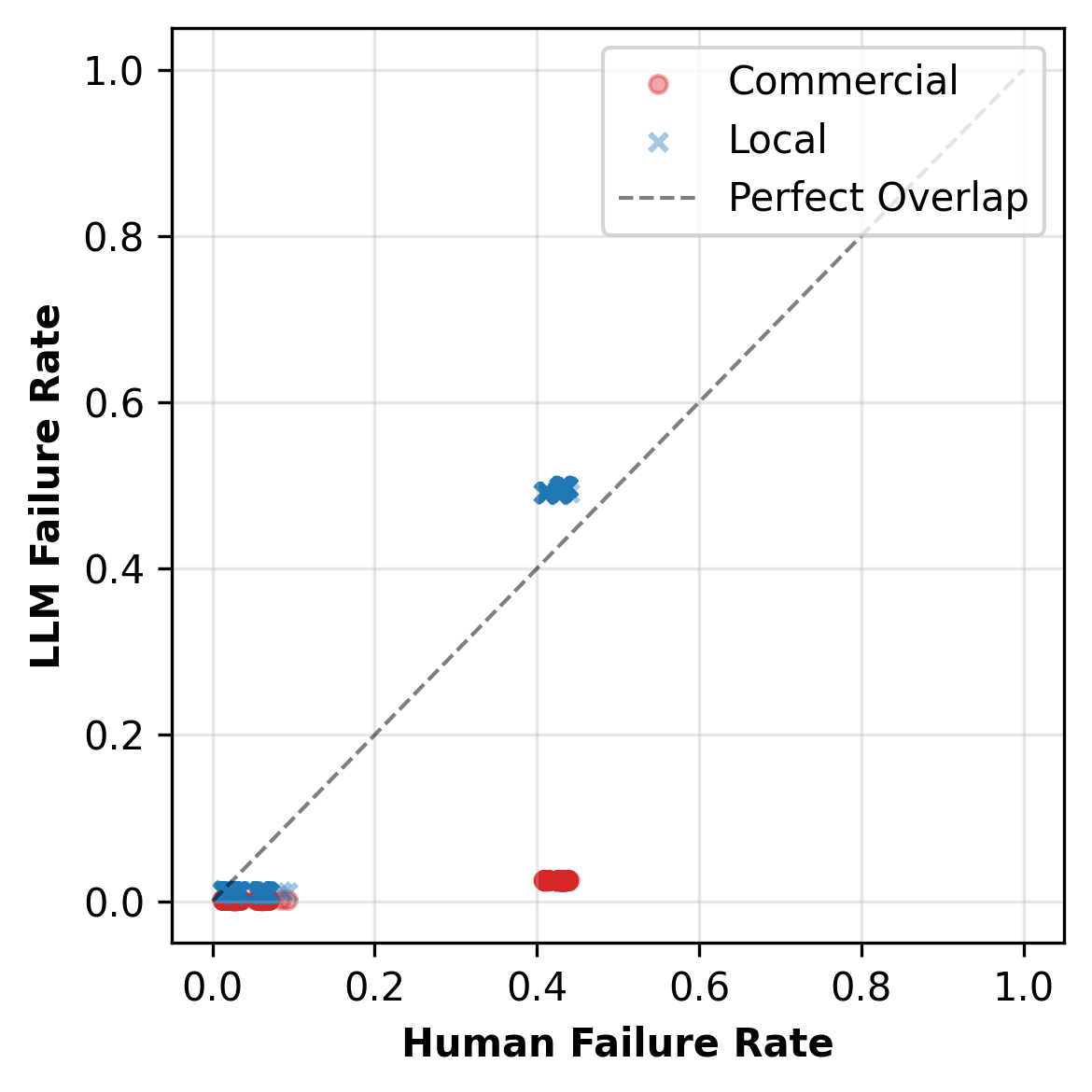}
        \centerline{(a) 3n+1}
    \end{minipage}\hfill
    \begin{minipage}{0.32\textwidth}
        \centering
        \includegraphics[width=\linewidth]{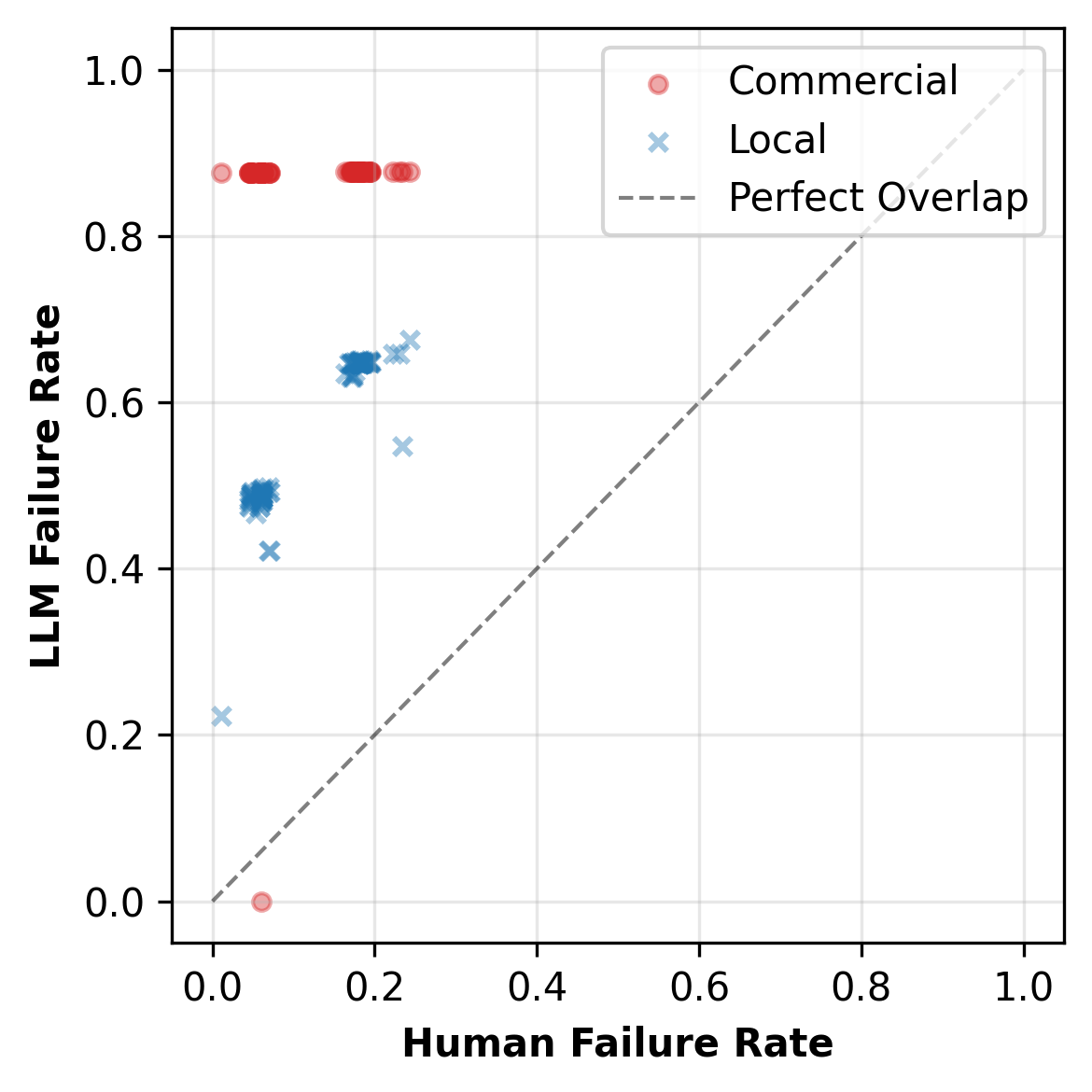}
        \centerline{(b) Factors and Factorials}
    \end{minipage}\hfill
    \begin{minipage}{0.32\textwidth}
        \centering
        \includegraphics[width=\linewidth]{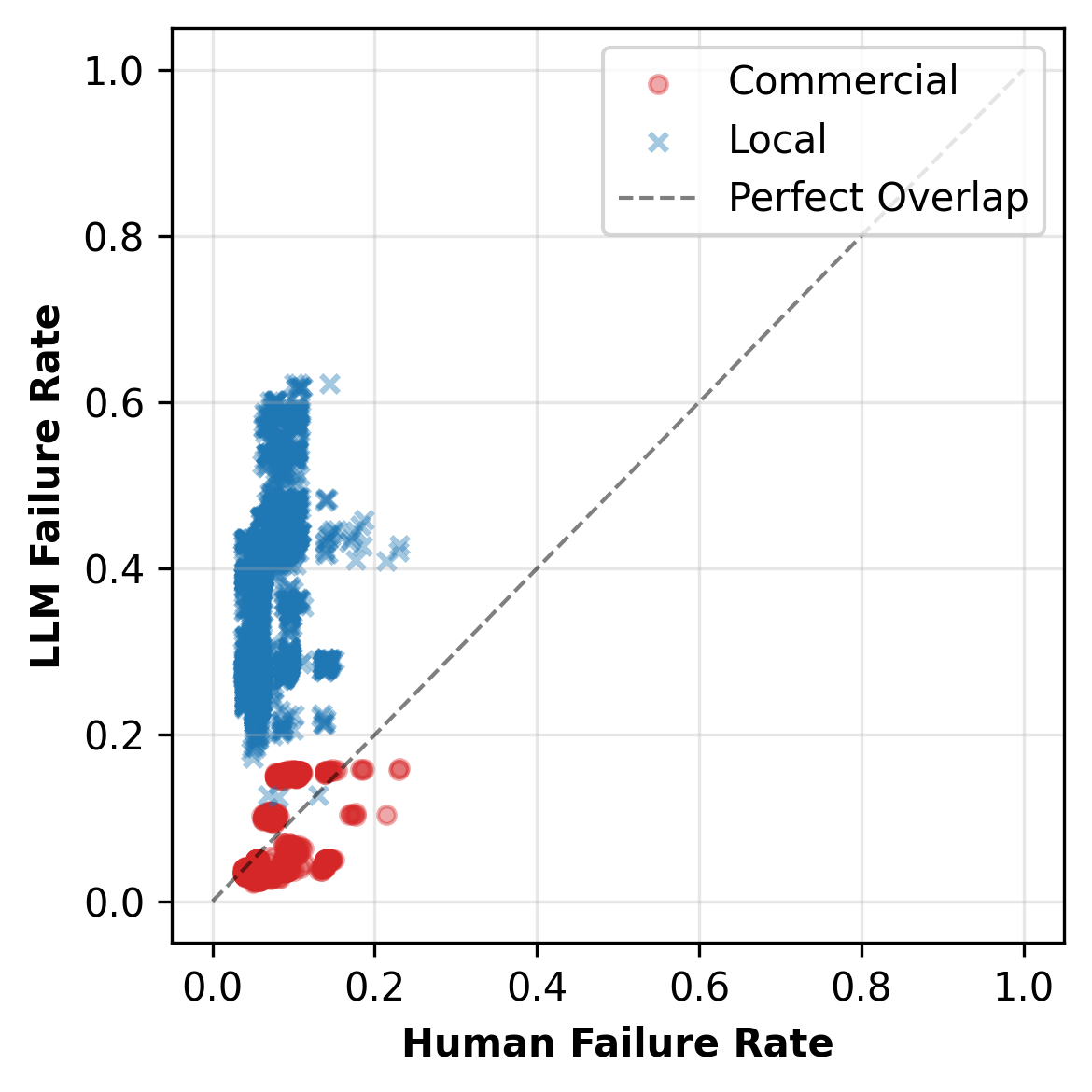}
        \centerline{(c) Factovisors}
    \end{minipage}
    \caption{Combined scatter plots comparing individual test case failure rates of human-written programs against Commercial and Local \gls{LLM} programs. Overlaid data demonstrates the degree of shared failures, ranging from tightly clustered (a) to dispersed (c).}
    \label{fig:combined_scatter_plots}
\end{figure*}

To determine whether \gls{LLM}-generated programs exhibit failure independence relative to human-written programs, we evaluated their failure profile overlap. Figure \ref{fig:combined_scatter_plots} maps this failure space using plots representing all test cases, as no single test case achieved a 0\% failure rate across all pools. The axes denote the probability of a test case failing in the human pool (x-axis) versus the \gls{LLM} pools (y-axis) (i.e., plotting 5,000 individual test cases for both \textit{3n+1} and \textit{Factovisors}, and 99 for \textit{Factors and Factorials}).

Empirical data demonstrates that failure overlap is highly specification-dependent. For \textit{3n+1}, failure profiles concentrate into tight clusters. While Local models strongly align with the human baseline, Commercial models display noticeable orthogonality: a significant cluster of 3,725 test cases (74.5\%) were systematically difficult for human developers ($\sim$43\% failure rate) but trivially resolved by Commercial \glspl{LLM} ($\sim$2.5\% failure rate). For \textit{Factors and Factorials}, structural overlap is negligible, with \gls{LLM} programs exhibiting uniformly high failure rates across all test cases. Finally, for \textit{Factovisors}, failure profiles show significantly higher dispersion than in \textit{3n+1}. This wide variance confirms that \glspl{LLM} and humans possess highly diverse failure patterns on a broad range of test cases, pointing to the effectiveness of cross-source heterogeneous pairing for reliability improvement.

\subsubsection{Effects of Generation Temperature}
In \glspl{LLM}, the temperature hyperparameter controls token creativity. In traditional software engineering, deterministic behavior is preferred; however, in \gls{NVP}, parameter-induced randomness may provide variance to mitigate common-mode failures.

To evaluate temperature effects on generated program failure rates and failure diversity, we grouped all valid \gls{LLM}-generated programs that supported temperature parameterization by generation temperature (0.50 to 1.50). We computed the diversity improvement factor for two configurations: homogeneous (pairing two \gls{LLM} programs generated at the same temperature) and heterogeneous (pairing an \gls{LLM} program generated at a specific temperature with a human program). 

Empirical data reveals a quantitative trade-off between baseline volume and failure diversity. As the generation temperature increases from 0.50 to 1.50, the volume of valid programs decreases across all specifications (e.g., dropping from 3,723 to 1,964 in \textit{Factovisors}), indicating that higher temperature values induce fatal syntax and semantic deviations.

However, for programs that successfully compile, higher temperatures yield measurable increases in homogeneous diversity. In both the \textit{3n+1} and \textit{Factovisors} specifications, the homogeneous improvement factor rises along the temperature scale, peaking at 3.51$\times$ and 4.73$\times$, respectively, at T=1.50. This confirms that modifying the generation temperature causes models to generate behaviorally diverse programs, reducing the probability of coincident failures compared to deterministic inferences. Conversely, the \textit{Factors and Factorials} specification shows relatively static homogeneous improvement ($\sim$1.15$\times$), demonstrating that temperature adjustments alone cannot overcome the pervasive baseline errors previously observed for this specification.

Crucially, while temperature effectively injects variance within the \gls{LLM} pool, it has no impact on heterogeneous diversity. Across all three specifications, the improvement factor derived from pairing an \gls{LLM} program with a human program remains constant regardless of the \gls{LLM}'s generation temperature ($\sim$2.34$\times$ for \textit{3n+1}, $\sim$8.20$\times$ for \textit{Factors and Factorials}, and $\sim$15.20$\times$ for \textit{Factovisors}). This indicates that while temperature alters the distribution of faults within the AI landscape, the fundamental orthogonal distance between \glspl{LLM} and human failures remains constant.

\section{Discussion}
\label{sec:discussion}

Our results show that \glspl{LLM} can offer a distinct, useful source of failure diversity. This section interprets the underlying causes of the observed empirical trends and discusses their implications for fault-tolerant system design.

\subsection{The Nature of LLM vs. Human Failures}
The \gls{EL}~\cite{eckhardt1985theoretical} model assumes that different developers will stumble on the same intrinsically difficult test cases. Our results confirm that \glspl{LLM} suffer from the same phenomenon, as evidenced by the homogeneous LLM reliability curves \jrc{(Figure \ref{fig:homogeneous_combined_reliability})} consistently remaining bounded by the independence line. Different AI models, even those with distinct parameter counts or architectures, share deep algorithmic biases, likely inherited from overlapping open-source pre-training corpora (e.g., GitHub, StackOverflow). 

However, while failures of \glspl{LLM}-based programs tend to be positively correlated, they tend to do so to a lesser extent than failures of human-written ones. As per the homogeneous analysis (Figure \ref{fig:homogeneous_combined_reliability}), \gls{LLM}-generated programs yielded higher reliability improvements than human programs, particularly for \textit{3n+1} and \textit{Factovisors} specifications.

Furthermore, the \textit{test case difficulty} for human programs and \glspl{LLM} ones are different. As seen in Figure \ref{fig:combined_scatter_plots}, while similar failure rates between Local LLMs and Humans are shown for the \textit{3n+1} specification, examples of disparate failure rates exist (such as with Commercial models in \textit{3n+1}, or across the other two specifications). Humans and AI models often struggle with different segments of the test suite. This contrast in \enquote{what makes a test case hard} is precisely what makes the cross-source combination of LLM and human code effective.

\subsection{An Example of Hybrid \gls{NVP} Effectiveness}
The historical barrier to the widespread adoption of \gls{NVP} has always been the prohibitive cost and time required to develop multiple independent implementations. LLMs fundamentally disrupt this economic constraint by enabling the rapid, low-cost generation of diverse programs. 

However, bearing in mind the need for baseline reliability, the most significant finding of this study is that this cost-effective diversity does not come at a penalty to fault tolerance; rather, it provides a distinct architectural advantage. By enforcing an \gls{LM} architecture that pairs one human-written program with one LLM-generated program, the empirical reliability improvement consistently surpassed the pure human baseline, the pure LLM baseline, and even the completely mixed (\gls{EL}) pool. This analysis provides a critical architectural insight for fault-tolerant system design: simply pooling human and AI outputs without enforcing structural heterogeneity is sub-optimal. To maximize the reliability gains obtainable from software diversity, system architects should seek to preserve meaningful heterogeneity between redundant components by drawing them from sufficiently different cognitive and computational sources, even though modern human-written code is increasingly shaped by some degree of AI assistance.

In some cases, this forced human-AI diversity even crossed the theoretical independence bound, demonstrating a possibility of negative failure correlation. The human and the AI effectively compensated for each other's respective weaknesses for these, relatively simple,  programs.

%\subsection{The Limits of \enquote{Free} Diversity}
%The results highlight the limitations of inducing diversity purely through hyperparameter tuning or automated feedback loops. 
Additionally, adjusting the generation temperature successfully increased the divesity effectiveness within the LLM pool (improving homogeneous LLM pairings), but it had no impact on the heterogeneous cross-source reliability. This implies that temperature alters the distribution of errors within the \enquote{LLM capability space}, but it does not bridge the fundamental cognitive gap between AI and human reasoning.

%Similarly, the iterative self-correction analysis (PMF evolution) demonstrated that while simple compiler feedback loops are excellent at repairing syntax faults (shifting partial failures toward perfection), they are largely impotent against fundamental algorithmic hallucinations. If a model completely misunderstands the mathematical premise of a specification in its 0-shot attempt, three rounds of automated error logs will rarely allow it to improve on $PFD=1.0$ state. True software diversity requires distinct logical paths, not simply randomly perturbed tokens or automated syntax patching.

\subsection{Language Forcing as a Catalyst}
Our results indicate that manipulating the constraints of the prompt, specifically by forcing the \gls{LLM} to generate code in different programming languages, acts as an effective catalyst for failure diversity. 

As observed in the heterogeneous language analysis (Figure \ref{fig:heterogeneous_combined_reliability}), pairing programs written in different languages (e.g., an \gls{LLM}-generated C++ program and a Java program) yielded significant reliability improvements, frequently pushing the system toward the independence bound and outperforming the effectiveness of language diversity observed for human-written programs. For instance, considering the PFD target of $5 \cdot 10^{-5}$ in the \textit{3n+1} specification, the enforced LLM C++ and Java pairing achieved an improvement factor of over 44,000x. This outperforms both the equivalent human heterogeneous pairing (192.8x) and a homogeneous LLM pair drawn from a unified pool (4,790x) at that exact same reliability threshold. This demonstrates that language forcing provides an order-of-magnitude leap in reliability improvement compared to the baseline variance achieved by simply randomizing temperature or model architecture within a single language.

These findings suggest that programming languages are not merely translation layers for \glspl{LLM}; they fundamentally alter the model's generative pathway. Consequently, the \gls{LLM} constructs the algorithmic logic differently, effectively sidestepping the common-mode failures that plague homogeneous language pools. While temperature scaling can offer minor intra-language variance, achieving structural diversity requires bounding models within distinct language ecosystems.

\subsection{Choosing Between Models: The Commercial vs. Local}
A final consideration in constructing \gls{LLM}-based fault-tolerant systems is the choice of model architecture and scale. Our results demonstrated a severe dichotomy in out-of-the-box reliability between Commercial APIs and Local (open-weight) models, which profoundly affects their viability as redundant components in a \gls{LM}~\cite{littlewood1989conceptual} architecture.

A naive approach to software diversity might suggest pairing a human developer with a smaller, highly quantized Local model to maximize architectural heterogeneity. However, the data convincingly illustrates the danger of improvement paradox in this scenario. In the \textit{3n+1} specification at a target PFD of $10^{-5}$, pairing a human with a Local model yielded an astronomical improvement factor of over 2.9 million. While mathematically correct, this value is deceptive: it occurs because the Local model's high failure rate creates random, orthogonal failure that rarely overlaps with the human's test cases failures. This results in a massive $R$ value, but the high reliability of the 1-out-of-2 system is almost entirely due to the human program alone.

Conversely, pairing humans with Commercial models resulted in more effective failure diversity. In the \textit{Factovisors} specification at $10^{-5}$ (not shown in our figures), the Human-Commercial pair achieved a 1,251.1x improvement compared to the Human-Local pair's 167.0x. We conjecture that programs generated by Commercial models fail in complementary ways to the ones written by humans.

%--------------------------------------------------------------------- 
\section{Threats to Validity}
\label{sec:threats}

\subsection{Internal Validity}
%A first threat concerns the comparability of the human-written and \gls{LLM}-generated populations. The two sets of variants were not produced under identical conditions: human-written solutions were drawn from the UVa ecosystem and filtered to retain the first valid submission per author, whereas \gls{LLM}-generated solutions were produced under a bounded iterative repair process with a fixed maximum number of retries. The resulting comparisons should therefore be interpreted as characterizing the reliability-oriented usefulness of \gls{LLM}-generated variants under a controlled generation setting, rather than as a direct equivalence between the two production processes.

Possible prior exposure of \glspl{LLM} to related specification statements or implementations may influence the results. Although the exact human-written corpus used here is not public in the analyzed form, similar competitive-programming specifications, programs, and discussions have long been available online. The results are thus best understood as characterizing failure overlap and reliability behavior in the evaluated outputs, rather than generation from a fully unseen specification space.

% The filtering criteria used to construct the diversity pools also affect interpretation. Excluding completely incorrect solutions \((\gls{PFD}=1)\) follows prior software-diversity studies and was necessary to build comparable pools for the reliability analysis. The reported results therefore characterize the retained non-trivial variants rather than the full set of attempted solutions.

Because the main \gls{LLM}-generated programs analysis focuses on the first retained program, additional retries could in principle improve some flawed programs and alter the composition of the diversity pools. To examine this, we evaluated the effect of iterative prompting (these results are omitted here due to space constraints), which indicates that iterative correction recovers some partial failures but has limited impact on completely incorrect programs, and does not materially change the overall diversity trends.

\subsection{Construct Validity}

A first construct-validity concern is whether the measures used in this study adequately capture software diversity for reliability improvement. The analysis represents each program by its failure vector over a test suite and summarizes behavior through \gls{PFD} and reliability-improvement curves in homogeneous and heterogeneous pairings. These measures are aligned with the classical software-diversity literature, but they do not capture all possible notions of diversity, such as stylistic, semantic, or maintainability-oriented variation. The focus here is explicitly on reliability-oriented diversity, for which failure diversity is of interest.

A second threat is that the composition of the test suite inherently defines the context over which diversity and reliability are evaluated. The resulting \gls{PFD} values and failure-overlap patterns are, therefore, conditional on the specific test cases selected. Different distributions of test cases or broader test suites could lead to different difficulty landscapes and reliability gains. To mitigate this, the study employs exhaustive or near-exhaustive test suites for the selected specification spaces and applies the exact same test suites to evaluate both human-written and \gls{LLM}-generated programs.

A third concern relates to the interpretation of \gls{LLM}-generated diversity. In this study, diversity is induced through model family, programming language, temperature, and iterative repair path. These are practical and controllable sources of variation, but they remain indirect proxies for deeper differences in logic, algorithm selection, and failure behavior. For this reason, the analysis is grounded in observed failure behavior rather than in syntactic variation alone.

\subsection{External Validity}

The study is based on three competitive-programming specifications and on historical UVa submissions. These tasks are well suited to large-scale empirical analysis because they provide many implementations of the same specification, but they do not represent industrial software or full software-development processes. 

Because the human-written corpus predates routine \gls{LLM} use, the cross-source analysis is not a direct model of current mixed human/\gls{LLM} development. At the same time, this corpus remains valuable as a rare large-scale pre-\gls{LLM} human baseline, enabling direct comparison between unaided human-written code and \gls{LLM}-generated code, and allowing us to assess whether \gls{LLM}-generated programs add useful diversity beyond that already present in human-written programs.

Due to time and cost constraints, the study does not cover the full space of commercial and proprietary \glspl{LLM}. Instead, it evaluates a broad set of open-weight models, together with some commercial ones, spanning multiple architectures, sizes, languages, and generation settings. While not exhaustive, this still provides a useful view of how different model families and generation regimes contribute to diversity effectiveness.

%--------------------------------------------------------------------- 
\section{Conclusion}
\label{sec:conclusion}

The ability of \glspl{LLM} to generate code on demand creates a new setting in which software diversity can be produced at scale, making it important to assess whether, and to what extent, these programs exhibit failure diversity. In this paper, we investigated software diversity for reliability improvement through \gls{LLM}-generated code. We compared historical human-written programs, large pools of \gls{LLM}-generated programs, and heterogeneous human–\gls{LLM} pairings, with emphasis on failure overlap and the reliability gains obtainable from redundancy. Our results show that \gls{LLM}-generated programs can yield measurable reliability gains, but that these gains depend strongly on the specification, programming language, and generation regime. They also show that \glspl{LLM} do not eliminate the classical challenge of correlated failures; rather, they provide a inexpensive source of heterogeneous programs whose value depends on how much they reduce failure overlap. These results suggest that \glspl{LLM} have the potential to support multi-version software by making it practical to systematically generate and evaluate heterogeneous programs whose diversity can be leveraged for reliability improvement.

Future work includes studying diversity in contemporary \gls{LLM}-assisted human development settings, extending the analysis to richer notions of structural or algorithmic diversity, including additional models and generation workflows, and evaluating whether similar patterns hold for larger and more realistic software artifacts.

\bibliographystyle{IEEEtran}
\bibliography{IEEEabrv,references}

\end{document}